\documentclass[fleqn,usenatbib]{mnras}
\usepackage{newtxtext,newtxmath}
\usepackage[T1]{fontenc}
\DeclareRobustCommand{\VAN}[3]{#2}
\let\VANthebibliography\thebibliography
\def\thebibliography{\DeclareRobustCommand{\VAN}[3]{##3}\VANthebibliography}

\usepackage{graphicx}
\usepackage{amsmath}	
\usepackage{orcidlink}
\usepackage{hyperref} 
\usepackage{breakurl} 

\def\grs{GRS\,1915+105}

\def\hei{He\,{\sc i}}
\def\heii{He\,{\sc ii}}
\def\ha{H$\alpha$}
\def\pfa{Pf$\alpha$}
\def\micron{\textmu m}
\def\p{$\pm$}
\def\ebv{$E_{\rm B-V}$}
\def\lbol{$L_{\rm Bol}$}

\def\nh{$n_{\rm H}$}
\def\nub{$\nu_{\rm b}$}
\def\av{$A_{\rm V}$}

\def\jwst{{\em JWST}}
\def\spitzer{{\em Spitzer}}

\def\mbh{$M_{\rm BH}$}
\def\Msun{M$_{\odot}$}

\def\gtsim{\mathrel{\hbox{\rlap{\hbox{\lower3pt\hbox{$\sim$}}}\hbox{$>$}}}}
\def\ltsim{\mathrel{\hbox{\rlap{\hbox{\lower3pt\hbox{$\sim$}}}\hbox{$<$}}}}
\def\nlyc{$Q({\rm H}^0)$}

\title[Mid-IR Spectral-Timing of \grs]{Rapid Mid-Infrared Spectral-Timing with \jwst:\\I. \grs\ during a MIR--bright and X-ray--obscured state\thanks{Dedicated to the memory of our colleague, Tomaso Belloni.}}
\author[JWST Timing Consortium]{\parbox{\textwidth}{
P.~Gandhi,\orcidlink{0000-0003-3105-2615}$^{1}$\thanks{E-mail: poshak.gandhi@soton.ac.uk (PG)}
E.~S.~Borowski,\orcidlink{0009-0002-6989-1019}$^{2}$ 
J.~Byrom,$^{1}$
R.~I. Hynes,\orcidlink{0000-0003-3318-0223}$^{2}$
T.~J. Maccarone,$^{3}$
A.~W. Shaw,$^{\orcidlink{0000-0002-8808-520X}}$$^{4}$
O.~K.~Adegoke,\orcidlink{0000-0002-5966-4210}$^{5}$
D.~Altamirano,$^{\orcidlink{0000-0002-3422-0074}}$$^{1}$ 
M.~C.~Baglio,$^{\orcidlink{0000-0003-1285-4057}}$$^{6}$
Y.~Bhargava,$^{\orcidlink{0000-0002-5967-8399}}$$^{7}$
C.~T.~Britt,$^{8}$
D.~A.~H.~Buckley,$^{\orcidlink{0000-0002-7004-9956}}$$^{9}$
D.~J.~K.~Buisson,$^{10}$
P.~Casella,\orcidlink{0000-0002-0752-3301}$^{11}$
N.~Castro Segura,$^{\orcidlink{0000-0002-5870-0443}}$$^{12}$
P.~A.~Charles,$^{1, 16}$
J.~M.~Corral-Santana,$^{\orcidlink{0000-0003-1038-9104}}$$^{13}$
V.~S.~Dhillon,$^{\orcidlink{0000-0003-4236-9642}}$$^{14, 15}$
R. Fender,$^{16}$
A.~G\'urpide,\orcidlink{0000-0002-2256-2704}$^{1}$
C. O. Heinke, $^{\orcidlink{0000-0003-3944-6109}}$$^{17}$
A.~B. Igl,\orcidlink{0000-0002-2844-5613}$^{2}$
C.~Knigge,$^{\orcidlink{0000-0002-8183-2970}}$$^{1}$
S.~Markoff,$^{\orcidlink{0000-0001-9564-0876}}$$^{18}$
G.~Mastroserio,\orcidlink{0000-0003-4216-7936}$^{19}$
M.~L.~McCollough,$^{\orcidlink{0000-0002-8384-3374}}$$^{20}$
M. Middleton,$^{\orcidlink{0000-0002-8183-2970}}$$^{1}$
J.~M. Miller,$^{\orcidlink{}}$$^{21}$
J. C. A. Miller-Jones,$^{\orcidlink{0000-0003-3124-2814}}$$^{22}$
S.~E. Motta\orcidlink{0000-0002-6154-5843}$^{23}$
J.~A. Paice,\orcidlink{0000-0003-1149-1741}$^{24}$
D.~D. Pawar,$^{25}$
R.~M. Plotkin,\orcidlink{0000-0002-7092-0326}$^{26}$
P. Pradhan,\orcidlink{0000-0002-1131-3059}$^{27}$
M.~E.~Ressler,\orcidlink{0000-0001-5644-8330}$^{28}$
D.~M.~Russell,\orcidlink{0000-0002-3500-631X}$^{29}$
T.~D.~Russell,$^{\orcidlink{0000-0002-7930-2276}}$$^{30}$
P.~Santos-Sanz,$^{\orcidlink{0000-0002-1123-983X}}$$^{31}$
T.~Shahbaz,\orcidlink{0000-0003-1331-5442}$^{15,32}$
G.~R.~Sivakoff,$^{\orcidlink{0000-0001-6682-916X}}$$^{17}$
D.~Steeghs,$^{\orcidlink{0000-0003-0771-4746}}$$^{12}$
A.~J.~Tetarenko,\orcidlink{0000-0003-3906-4354}$^{33}$
J.~A.~Tomsick,\orcidlink{0000-0001-5506-9855} $^{34}$
F.~M. Vincentelli,\orcidlink{0000-0002-1481-1870}$^{1}$
M. George,\orcidlink{0009-0009-7908-4908}$^{1}$
M.~Gurwell,$\orcidlink{0000-0003-0685-3621}$$^{20}$
R.~Rao$\orcidlink{0000-0002-1407-7944}$$^{20}$
\begin{center} (JWST Timing Consortium) \end{center}
\begin{center} {\small \textit{Affiliations can be found at the end}}\end{center}
}
}

\date{Received in original form 2024 Jun 21}
\pubyear{\the\year{}}

\begin{document}
\label{firstpage}
\pagerange{\pageref{firstpage}--\pageref{lastpage}}
\maketitle
\begin{abstract}We present mid-infrared (MIR) spectral-timing measurements of the prototypical Galactic microquasar \grs. The source was observed with the Mid-Infrared Instrument (MIRI) onboard \jwst\ in June 2023 at a MIR luminosity $L_{\rm MIR}$\,$\approx$\,10$^{36}$\,erg\,s$^{-1}$ exceeding past IR levels by about a factor of 10. In contrast, the X-ray flux is much fainter than the historical average, in the source's now--persistent `obscured' state. The MIRI low-resolution spectrum shows a plethora of emission lines, the strongest of which are consistent with recombination in the hydrogen Pfund (Pf) series and higher. Low amplitude ($\sim$\,1\%) but highly significant peak-to-peak photometric variability is found on timescales of $\sim$\,1,000\,s. The brightest Pf\,(6--5) emission line lags the continuum. Though difficult to constrain accurately, this lag is commensurate with light-travel timescales across the outer accretion disc or with expected recombination timescales inferred from emission line diagnostics. Using the emission line as a bolometric indicator suggests a moderate ($\sim$\,5--30\% Eddington) intrinsic accretion rate. 
Multiwavelength monitoring shows that \jwst\ caught the source close in-time to unprecedentedly bright MIR and radio long-term flaring. Assuming a thermal bremsstrahlung origin for the MIRI continuum suggests an unsustainably high mass-loss rate during this time unless the wind remains bound, though other possible origins cannot be ruled out. PAH features previously detected with \spitzer\ are now less clear in the MIRI data, arguing for possible destruction of dust in the interim. These results provide a preview of new parameter space for exploring MIR spectral-timing in XRBs and other variable cosmic sources on rapid timescales.
\end{abstract}

\begin{keywords}
X-ray binaries -- infrared -- time-series\end{keywords}



\section{Introduction}
\label{sec:intro}

X-ray binaries (XRBs) host a compact object, either a black hole or a neutron star, in a bound orbit with a secondary star from which matter is accreted. XRBs represent the endpoints of massive stellar evolution in binaries. Their compact sizes also make excellent laboratories for studying accretion as a rapid time-domain phenomenon, and there is now a rich heritage of observations studying their flux variations across a range of timescales \citep[e.g., ][ and references therein]{belloni10}. Variability is also an important tool for disentangling emission from multiple physical components (e.g. the accretion disc, corona, jet, donor star; \citealt{vanderklis95, markoff01, Gandhi-2011, Veledina-2013, malzac14} and references therein) that can overlap spectrally, and make studies of XRBs challenging.

\grs\ is the prototypical \lq microquasar', displaying apparently superluminal relativistic jets and prolific X-ray variability patterns \citep{mirabel94, belloni00}. Continuum variability spanning a wide range of timescales has been observed and found to be correlated across many wavelengths, interpreted as evidence for disc-jet coupling and plasma ejections on timescales of $\sim$\,30\,min and longer \citep{fender97, Mirabel-1998, Eikenberry-1998, rothstein05,neilsenlee09,ueda10,vincentelli23}. With an integrated Galactic reddening \ebv\,=\,8.9\,$\pm$\,0.7 mag along its line-of-sight \citep{schlafly11}, the source is too extincted for most optical observations. But sub-minute near-infrared continuum variability has also been observed to be correlated with X-rays \citep{Lasso-Cabrera-2013}. 

The source has been observed in the mid-infrared (MIR) with previous missions. \citet{rahoui10} reported a detailed analysis of \spitzer\ spectral observations, finding signatures of dust together with an irradiated accretion disc, with both components responding to the intrinsic X-ray variations over long timescales of months (but see \citealt{harrison14}). Though no short-term emission line variability study was reported from the \spitzer\ observations (perhaps because the integration times, typically about 30 minutes, did not sample the variability well), evidence for such variations has been reported in the near-infrared by \cite{eikenberry98_spec}, who identified radiative pumping as the cause of common emission line and continuum flux trends. 

Since 2018, the source has also entered a new and puzzling spectral state, with the X-ray flux having systematically faded by a factor of 10--100 compared to its historical persistent average. Nonetheless, there are multiple indications -- from X-rays and radio monitoring, together with detection of near-IR winds -- that the source remains active and intrinsically luminous, with the faintness likely resulting from enhanced obscuration \citep{negoro18,miller20, balakrishnan21, motta21, sanchezsierras23}. The cause and nature of this `obscured' state remain unclear, though a change in the outer disc geometry, triggering of multiwavelength winds, and/or a change in accretion rate have been proposed \citep{miller20, neilsen20, balakrishnan21, sanchezsierras23}. 

\begin{figure*}
    \centering
    \hspace*{-1.2cm}
    \includegraphics[angle=0,width=1.06\textwidth]{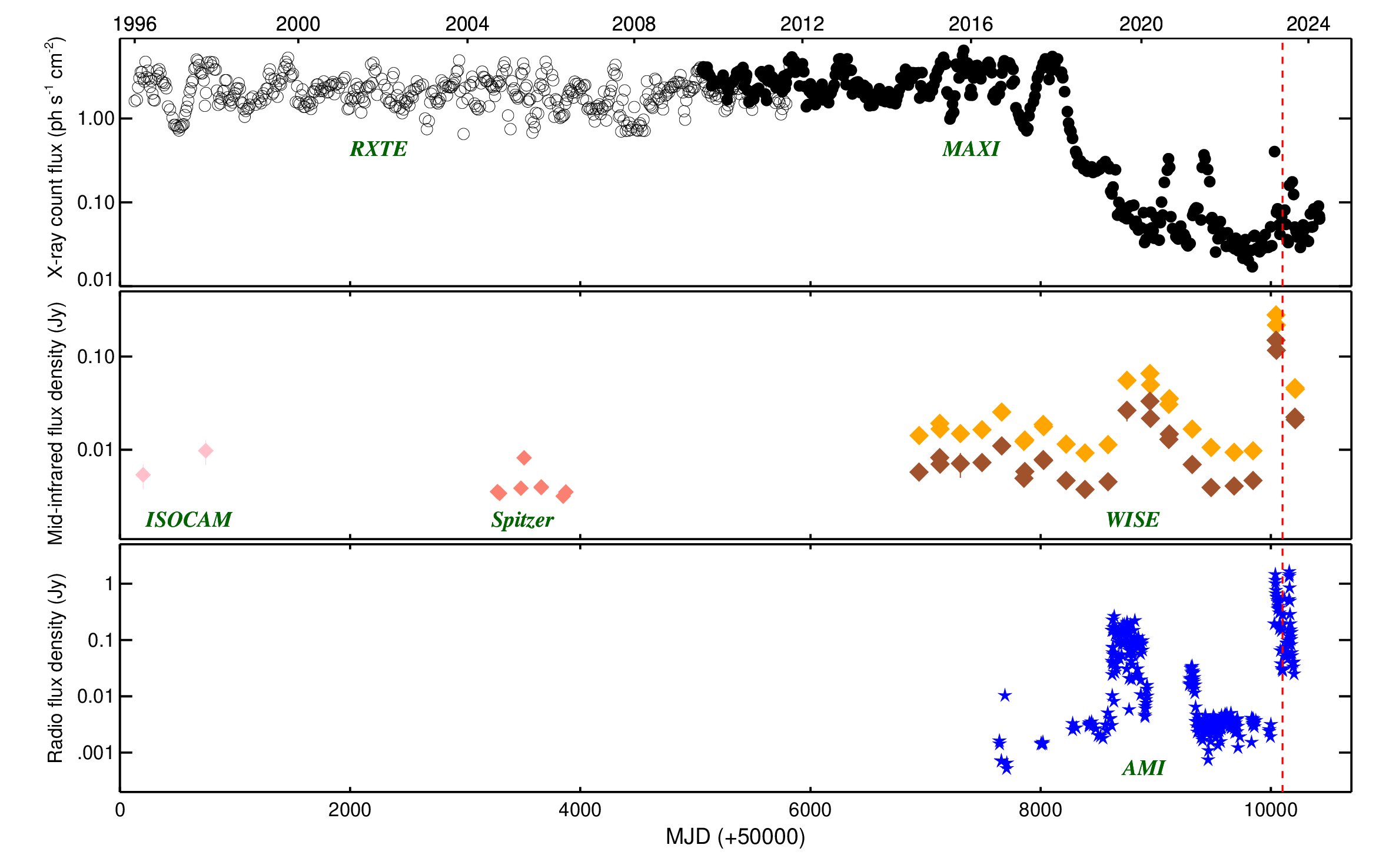}
        \hspace*{-1cm}
    \caption{{\em (Top)} Long-term X-ray light curve monitoring of \grs\ with MAXI (2--20\,keV; filled circles) and {\em RXTE} ASM (1.5--12\,keV; unfilled circles). The {\em RXTE} count rates have been normalised to match the MAXI data at their peaks, simply in order to plot and compare them simultaneously. {\em (Middle)} Long-term MIR flux densities from ISOCAM LW2 (5--8.5\,\micron; 1996--1997), \spitzer\ IRS 5--5.5\,\micron\ (2004--2006), and NEOWISE W1 (orange diamonds; shifted by 2$\times$ for display purposes) and W2 (brown) light curves from 2014 onwards. {\em (Bottom)} Long-term Arcminute Microkelvin Imager - Long Array (AMI-LA) radio light curve. Data are taken at 15\,GHz. The \jwst\ observation date (2023 Jun 06) is indicated by a vertical red line.}
    \label{fig:longterm}
\end{figure*}

Fig.\,\ref{fig:longterm} shows the long-term multiwavelength behaviour of the source, including X-rays, radio and MIR monitoring data. X-ray all-sky monitors \citep{rxte, rxteasm, maxi} show the systematic flux decline post-2018, with later emission dominated by a few flares reaching at most a factor of a few below previous fluxes. The middle panel shows long-term MIR measurements (in bands close together in wavelength centred around 3.4 to 8.5\,\micron) from the {\em ISO} mission \citep{iso, isocam, isocamlws}, \spitzer\ IRS \citep{spitzer, spitzerirs}, and NEOWISE observations \citep{wise, neowise}. Finally, the bottom panel shows AMI-LA \citep{ami} 15\,GHz radio monitoring since 2016. Both the radio and the MIR light curves show pronounced long-term flaring in recent years. Moreover, the NEOWISE data show flares during the obscured state reaching an order-of-magnitude brighter than historical MIR fluxes during the X-ray bright state -- all indicators of an intrinsically active source. 

\begin{table}
    \centering
    \caption{Physical parameters of \grs\ adopted in this work}
    \begin{tabular}{l|r}
    \hline
    Distance $d$      &  9.4\,\p\,1.0~kpc\\
    Orbital period $P$      &  33.85\,\p\,0.16~d\\
    Black hole mass \mbh & 11.2\,\p\,1.7~\Msun \\ 
    Inclination $i$ & 64\,\p\,4~deg \\
    Mass ratio $q (=M_2/M_1)$  & 0.042\,\p\,0.024 \\
    Binary separation $a$ & 232.2\,\p\,12.0~lt-sec\\
    Roche Lobe $r_L/a$ & 0.16\,\p\,0.03\\
    Extinction $A_{\rm V}$ & 19.2\,\p\,0.17 mag\\
    \hline
    \end{tabular}
    \label{tab:pars}
\end{table}

Here, we present \jwst\ mid-infrared (MIR) spectral-timing of \grs. A \jwst\ observation was conducted close in-time to multiwavelength long-term flaring, which included an event of exceptional brightening in the radio and MIR (Fig.\,\ref{fig:longterm}), presenting a unique opportunity to study the source in the MIR during its novel X-ray obscured state. We aim to try and understand the origin of the MIR emission and shed light on the circumnuclear environment in this new state. 

\jwst\ was not designed as a dedicated timing observatory. Nonetheless, its capabilities allow interesting new timing parameter space to be exploited, which we demonstrate here to study rapid (stochastic) accretion variability as a use case. This is the first in a series of papers dedicated to exploring MIR spectral-timing for XRBs. The sensitivity of these data are orders-of-magnitude superior to previous MIR observations of the source, offering new possibilities in precision spectral-timing exploration.  

A kinematic distance of 9.4\,\p\,0.6\,(stat.)\,\p\,0.8\,(sys.)\,kpc to \grs\ has been estimated \citep{reid23}, which is consistent with a radio parallax distance of 8.6$^{+2.0}_{-1.6}$\,kpc and yields a black hole mass of \mbh\,=\,11.2\,\Msun\ and a jet inclination of $i$\,=\,64\,\p\,4\,deg \citep{reid14}. The secondary is an evolved K1/5\,III star in a long orbit $P$\,=\,33.85\,\p\,0.16\,days \citep{greiner01, steeghs13}. These and other adopted source parameters from \citet{casaresjonker14} are listed in Table\,\ref{tab:pars}. The adopted value of optical extinction from \citet{chapuis04} will be discussed later. 

\section{Observations}
\label{sec:obs}

\grs\ was observed on 2023 June 06 using MIRI \citep{miri} in the Slitless Low Resolution Spectroscopy (LRS) mode \citep{kendrew15}. This was a time-series observation with a continuous stare on-target through the SLITLESS PRISM sub-array, with no dithering. The first on-source science frame started at UT\,08:57:49.86. A separate work will examine the absolute time accuracy of the mission clock (Shaw et al., in prep.). Observations were performed in the standard non-destructive FASTR1 mode (cf. \citealt{ressler15, dyrek24}), with a series of individual frames (or `groups'), each 0.159\,s in duration, combining charge on a `ramp' to yield a cumulative `integration'. A total of nINT\,=\,4,150 integrations were obtained, each comprising nGRP\,=\,10 science groups. The detector is reset at the end of each integration, resulting in a dead time of one group between successive integrations. The time of the final science frame was UT\,10:58:48.44, yielding a full science observation of just over 2\,hours. 

The data were processed through \jwst\ pipeline version 1.13.3 (build 10.1). The pipeline itself comprises an elaborate set of intertwined stages and is still in the early (post-launch) stages of development. Its main stages are: (1) detector level calibration, (2) spectroscopic calibration, and (3) outlier detection and spectral extraction. 

A number of inadequacies were identified in our run of the pipeline. Stage 1 was found to be over-flagging pixels as bad. This is likely to be a result of the known `brighter-fatter' effect of the MIRI detectors \citep{argyriou23} and its handling in the current version of the pipeline (STScI, priv. comm.). In consultation with the instrument team, we reprocessed the data through the pipeline by setting the {\tt flag\_4\_neighbors} flag to be {\tt false} and the {\tt JUMP rejection\_threshold} to 15\,$\sigma$ (from its default value of 4). Stage 3 of the pipeline was also found to be introducing substantial scatter between integrations in its outlier detection step, especially within the first few integrations when there is also a known detector settling delay. Again, in consultation with STScI and while this issue is investigated, we chose to forego the outlier detection step entirely. This ends up leaving a few residual artefacts that are not filtered out by the pipeline and need to be individually accounted for in later analysis (cf., \S\,\ref{sec:timingresults}), but should not impact the bulk of our analysis. 

The spectral trace across all integrations was examined by eye and was found to be stable at the sub-pixel level throughout the observation. A rectangular aperture of width 5 pixels (0.55 arcsec) was used for spectral extraction. The pipeline automatically accounts for wavelength-dependent aperture flux losses. Background subtraction is not implemented by default for slitless observations. We used two rectangular source-free apertures (experimenting with 8--11\,pixel widths) for computing and subtracting the background in each integration. During analysis of the source r.m.s variability (discussed in \S\,\ref{sec:results}), we found that the pipeline does not incorporate background uncertainties in its total extracted flux error, even when background subtraction is implemented (confirmed by STScI; priv.\,comm.). We attempted a manual correction by extracting background flux errors and adding them in quadrature to the source flux errors, but we suspect that some uncertainties are still being underestimated at the red end of the MIRI wavelength range. This is discussed in Appendix\,\ref{sec:1033correction} and could be related to known issues with read-noise subtraction, which should increasingly dominate the error budget at long wavelengths.\footnote{\url{https://jwst-docs.stsci.edu/depreciated-jdox-articles/jwst-calibration-pipeline-caveats/jwst-miri-lrs-pipeline-caveats}} The bulk of our analysis herein should not be strongly impacted by these factors, except at the longest wavelengths which we highlight in relevant places in the text later on. 

We extract spectra across the full delivered wavelength range of 4.5--14\,\micron, though it should be noted that the primary calibrations for MIRI are well understood up to 12\,\micron, and there remain uncertainties at longer wavelengths. Our data also become strongly background-dominated above $\approx$\,13\,\micron. In order to check the spectral calibration, we conducted a comparison of MIRI data obtained in an identical (LRS SLITLESS PRISM TSO) mode as part of a commissioning observation. These data, of the star L168--9b, showed very similar spectral features at the two wavelength ends of the spectrum, a clear indication of imperfect spectral calibration. We used these commissioning observations to obtain a wavelength-dependent spectral calibration correction, which was then applied to \grs. The details of this procedure can be found in Appendix\,\ref{sec:1033correction}. The correction factor introduces only small changes ($\approx$ a few per cent) over the bulk of the primary MIRI range, but becomes more important above 12\,\micron. 

Finally, we note that the pipeline delivers source signal at wavelengths extending down to $\lambda$\,$\approx$\,4\,\micron. However, there is a known spectral fold-over below $\approx$\,4.5\,\micron\ in slitless mode, so reflection and scattering causes longer wavelength light to contaminate detector regions characteristic of shorter wavelengths \citep{miricommissioning}. We thus restrict our analysis to $\lambda$\,$>$\,4.5\,\micron.

\section{Results}
\label{sec:results}

\subsection{Spectroscopy}
\label{sec:specresults}

\begin{figure*}
    \centering
    \includegraphics[angle=0,width=\textwidth]{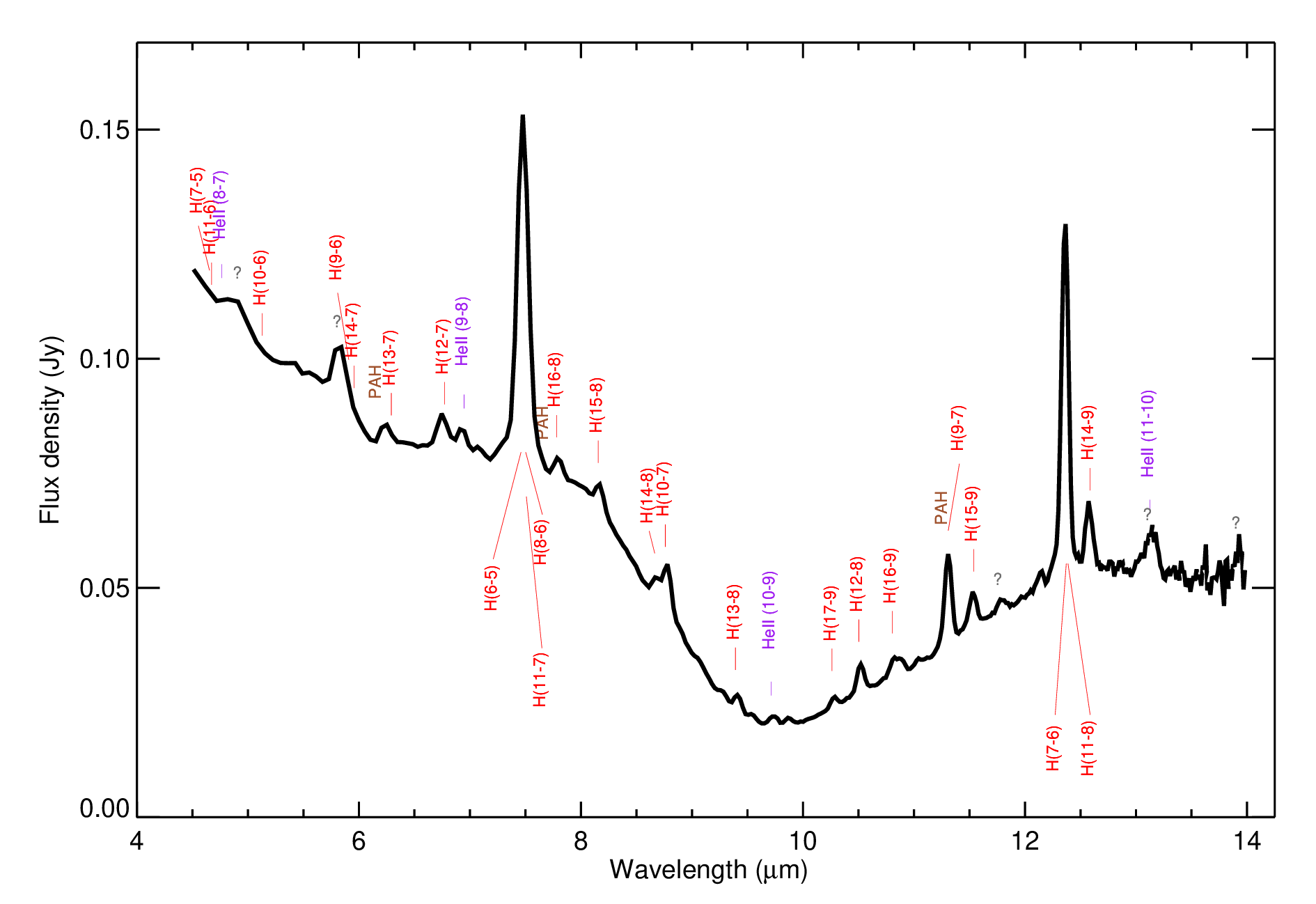}
    \caption{MIRI/LRS spectrum of \grs. The annotated identifications are not meant to be complete. Furthermore, while some potential line identifications are listed, the low resolution precludes unambiguous identification in many cases.}
    \label{fig:spec}
\end{figure*}

Fig.\,\ref{fig:spec} shows the final average MIRI LRS spectrum of \grs\ across the entire wavelength range of 
4.5--14\,\textmu m. A bright continuum is detected, peaking above a flux density $F_\nu$\,$\approx$\,100\,mJy at the short wavelength end ($\approx$\,5\,\micron). A deep absorption feature is detected over $\approx$\,9--11\,\micron, which can be attributed to Silicates (see \S\,\ref{sec:discussion}).

Many emission lines are detected, the most prominent of which are hydrogen recombination features from the Pfund (Pf; $n_l$\,=\,5) and Humphrey (Hp; $n_l$\,=\,6) series.\footnote{Appendix\,\ref{sec:hseries} presents a reference list of some of the key H and \heii\ emission lines that lie in the MIRI range.} Potential identifications of many features are annotated in the figure. Several of these are strong and significant, given the high signal-to-noise ratio (S/N) of the data. But in many instances, exact identification of features remains ambiguous, primarily due to the low spectral resolution ($R$) of the observing mode which ranges from $R$\,$\approx$\,40 ($\Delta \lambda$\,=\,0.125\,\micron; $\Delta v$\,=\,7,500\,km\,s$^{-1}$) near 5\,\micron\ to $\approx$\,160 ($\Delta \lambda$\,=\,0.063\,\micron; $\Delta v$\,=\,2,000\,km\,s$^{-1}$) at 10\,\micron\ \citep{kendrew15}. 

Interpretations of the rich emission line spectrum and the continuum will be discussed shortly. Detailed line investigation (especially accounting for blending) is, however, beyond the scope of the present work, and is likely to require higher spectral-resolution observations. Here, we present the results of basic line fitting to a few specific emission lines that will be discussed below. 

Profile fits were carried out assuming a single Gaussian component atop a local linear (first-order) continuum. This allowed measurements of centroid wavelengths, full-widths at half maximum (FWHM) and dereddened\footnote{Dereddening corrections are detailed in \S\,\ref{sec:powerdiscussion}.} fluxes. These results can be found in Table\,\ref{tab:emlines}.

There is no consistent pattern of centroid wavelength shifts from the most prominent expected line identifications, with the main uncertainty being possible confusion related to line blending. Regarding line widths, the measured FWHM range over\,$\approx$\,0.073--0.115\,\micron. The latter value is for the strongest feature peaking around $\lambda$\,=\,7.48\,micron and consistent with H\,Pf(6--5). At this wavelength, the expected line width for an unresolved line is FWHM\,=\,0.075\,micron ($R$\,$\approx$\,99.5) based upon a linear fit to the spectral resolution trace of \citet{kendrew15}. Thus, this feature appears to be significantly resolved. If this can be attributed solely to a single emission line, the implied velocity width is $\Delta v$\,$\approx$\,3,500\,km\,s$^{-1}$. By comparison, emission lines are broad as 2,100\,km\,s$^{-1}$ have been recently observed in the {\em NIR} in this source \citep{sanchezsierras23}. The main systematic uncertainty here remains our inability to deblend the various possible emission lines that may overlap and artificially broaden single-component fits.

\begin{table*}
    \centering
    \caption{Strong emission lines identified or discussed in the text body \label{tab:emlines}}
    \begin{tabular}{lccr}
    \hline
    \hline
       Wavelength of fit & FWHM     & Dereddened flux & Potential\\
    ($\mu$m)       & ($\mu$m) & ($\times 10^{-13}$ erg\,s$^{-1}$\,cm$^{-2}$) & identification\\
    \hline
        6.239\,\p\,0.004 & 0.096\,\p\,0.009 & 0.09\,\p\,0.04 & PAH feature $\sim$\,6.2\\
        \hspace*{0.9cm}$''$ & & & H\,(13--7)\,6.2919\\
        7.4774\,\p\,0.0003 & 0.115$\pm0.001$ & 10.03$\pm$0.03 & H\,Pf(6--5) 7.4599 \\
        \hspace*{0.9cm}$''$& & & H\,Hp(8--6) 7.5025\\
        \hspace*{0.9cm}$''$& & & H\,(11--7) 7.5081\\
        7.800\,\p\,0.005 & 0.08\,\p\,0.01 & 0.08\,\p\,0.01 & PAH feature $\sim$\,7.7\\
        \hspace*{0.9cm}$''$ & & & H\,(16--8)\,7.7804\\
        9.737$\pm$0.005 & 0.098$^{+0.008}_{-0.006}$ & 0.18$\pm$0.04 & \heii\,(10--9)\,9.7135 \\
        11.307$\pm$0.001 &  0.073$\pm$0.001& 0.62$\pm$0.03 & H\,(9--7)\,11.3087\\ 
        \hspace*{0.9cm}$''$ & & & \hei\ 11.3010\\
        \hspace*{0.9cm}$''$ & & & PAH feature $\sim$\,11.3\\
        12.3602$\pm$0.0006 & 0.074$\pm$0.001& 3.56$\pm$0.09 & H\,Hp(7--6) 12.3719 \\
        \hspace*{0.9cm}$''$& & & H\,(11--8) 12.3871\\
       \hline
       \hline
    \end{tabular}
    ~\\
    Fits based on single, symmetric Gaussian model parameterisation of the lines. Only the most prominent lines and those discussed specifically in the text are listed here. Ambiguous possible line identifications (which may not be complete) are listed in the final column, with wavelengths stated in \micron.
\end{table*}

\subsection{Timing}
\label{sec:timingresults}

\subsubsection{General Properties}

\begin{figure}
    \centering
    \includegraphics[width=1.01\columnwidth]{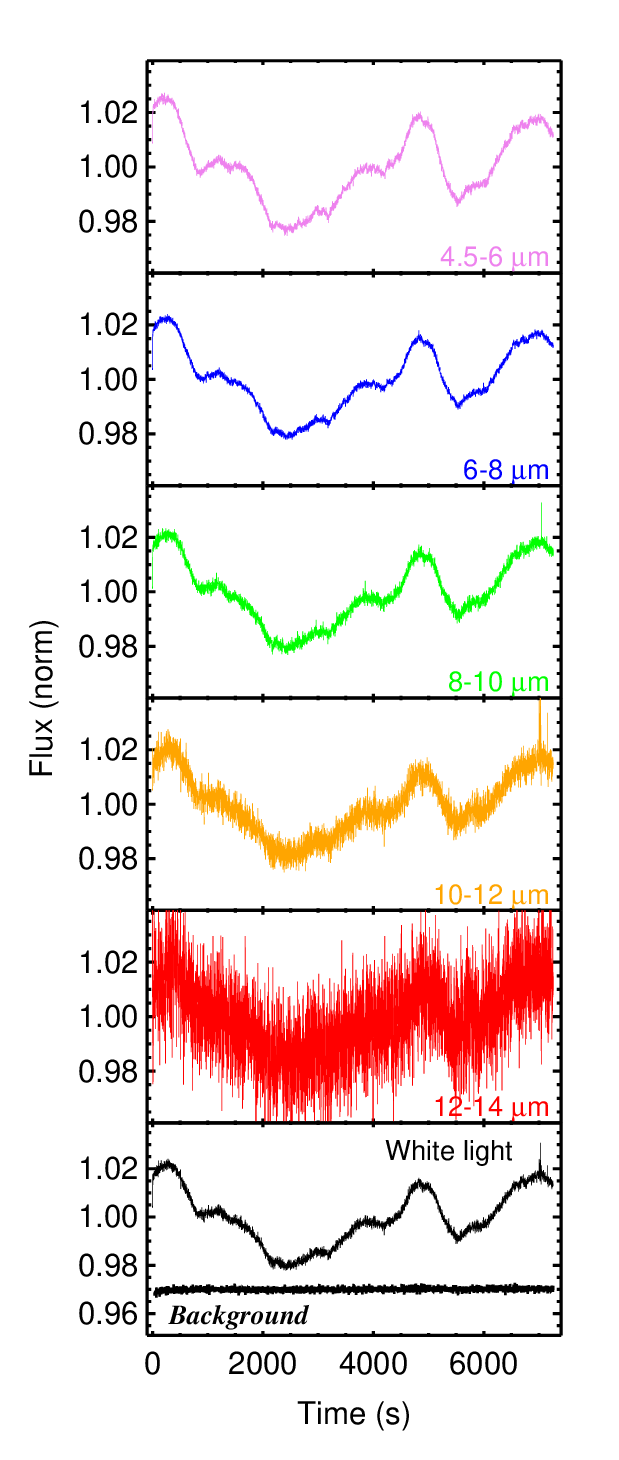}
    \vspace*{-0.5cm}
    \caption{Multi-wavelength time-series of all 4,150 integrations, split in annotated wavelength bins and normalised to the mean in each case. The spikes in the initial few integrations and also around a time of 7,000\,s are artificial features that have not been filtered out by the pipeline and should be ignored (they do not affect our lag analysis). The enhanced noise in the red filters is immediately apparent and is also an artefact of the pipeline (discussed in \S\,\ref{sec:obs} and in Appendix\,\ref{sec:1033correction}). The bottom panel reports the \lq white light\rq\ curve accumulated by the pipeline over 4.5--12\,\micron. The background light curve in the same band and units is also plotted, with an arbitrary offset applied on the y-axis in order to fit within the  panel.}
    \label{fig:lc}
\end{figure}

\begin{figure}
    \centering
    \includegraphics[angle=0,width=\columnwidth]{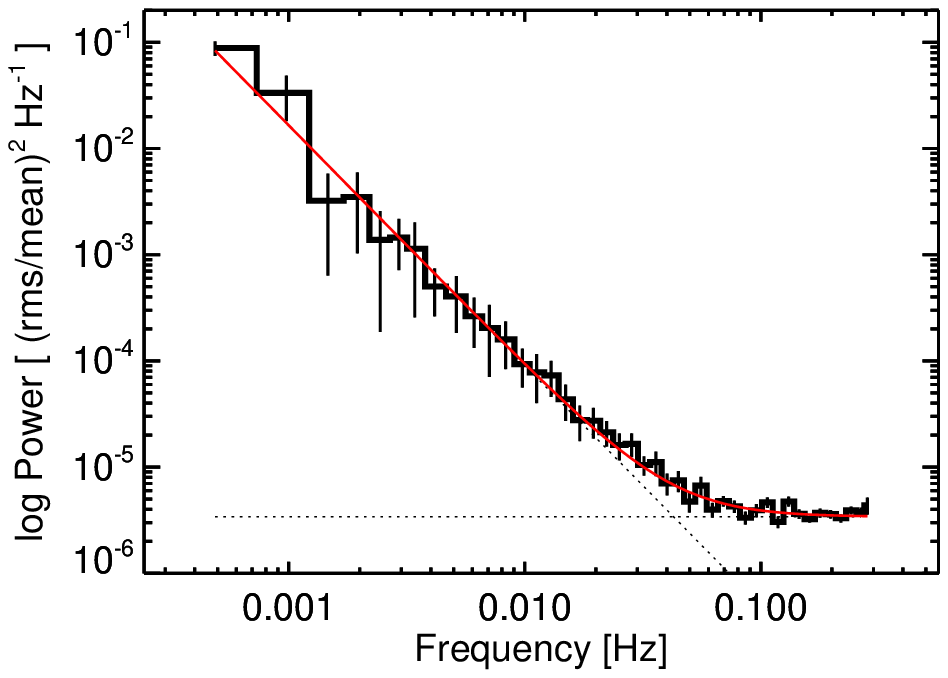}
    \caption{Power spectrum in white light, with the best-fit power-law of slope --2.3 overplotted, together with the constant white noise level. 
    }
    \label{fig:psd}
\end{figure}

The desire for timing analysis motivated our original choice of the MIRI slitless LRS mode \citep{kendrews18tso} for this observation. The data are collated in three-dimensional `cubes' comprising two-dimensional spectral images sampled at each integration. As a starting point to investigate the timing properties, the spectral-timing cubes can be summed over the full wavelength range to produce light curves in desired wavelength ranges. Fig.\,\ref{fig:lc} shows such light curves spanning short to long wavelength ranges, together with a `white-light' time series summed over the full nominal MIRI range of 4.5--12\,micron. 

As a result of the high source brightness, the typical cumulative signal-to-noise (S/N) ratio in each 1.59\,s integration of the white-light data is $\approx$\,700. This allows fine sensitivity to temporal flux variations, revealing weak ($\sim$\,per cent level) but significant fluctuations. A few obvious, but minor, artefacts remain in the light curve (e.g. the spike in the initial integrations and the cosmic rays near 7,000\,s), and are related to pipeline shortcomings (cf\,\ref{sec:obs}). These do not impact our inferences below. 

For the white-light data, we find an excess r.m.s of 1.1\% above statistical errors, defined as:

\begin{equation}
{\rm r.m.s._\lambda}=\frac{\sqrt{var_\lambda-\overline{{\sigma_{\rm err(\lambda)}^2}}}}{\overline{f_\lambda}}
\label{eq:rms}
\end{equation}

\noindent
where $\overline{f}$ denotes the mean flux of the light curve, $var$ is its raw variance, $\sigma_{\rm err}$ the individual errors measurements and $\overline{\sigma_{\rm err}^2}$ the mean square error \citep[cf., ][]{vaughan03}. All quantities above are functions of wavelength $\lambda$. 

The strongest (peak-to-peak) variations occur on timescales of $\sim$\,1,000--2,000\,s, and there are no obvious characteristic faster timescales. This was confirmed by computing the power spectral density (PSD) of the white-light time series, displayed in Fig.\,\ref{fig:psd}. This represents the squared modulus of the complex-valued Fourier spectrum of the light curve, and is normalised according to the r.m.s. of the light curve \citep[cf. ][]{belloni90, vaughan03}. The PSD was accumulated in three time segments, each 2,048\,s in length, which were then averaged to compute the mean and scatter. A light binning by a factor of 1.1 in logarithmic frequency bins has been applied. The PSD peaks at low frequencies, showing a steep fall from frequencies of $\approx$\,0.0005 to $\approx$\,0.05\,Hz, above which white noise starts to dominate. The effect of deadtime has not been corrected for in this PSD analysis, but its impact should be minimal given the short duration of the gap between integrations and the weakness of any rapid variations. The slope of the steep component is consistent with a simple red-noise power-law $P_\nu$\,$\propto$\,$\nu^{-\alpha}$, with $\alpha$\,=\,--2.3\,\p\,0.05. 

Investigating the typical variability as a function of wavelength, it can be seen from Fig.\,\ref{fig:lc} that the mean trace of the light curves in all wavelength ranges is very similar in terms of shape and peak-to-peak variability. However, the longer wavelengths above 10\,\micron\ also show substantial excess scatter around the instantaneous mean. This was confirmed through a wavelength-resolved r.m.s plot which also showed the same trend. This rise in variability at the longer wavelengths is a suspected artefact of the aforementioned underestimation of uncertainties at the red end (cf.\,\S\,\ref{sec:obs}), and we do not consider it to be real. 

At the wavelengths of many of the strong emission lines, the r.m.s. was found to be depressed. An exception in this regard was \heii\,(10--9)$\lambda$9.7\,\micron, where the r.m.s. was {\em enhanced} relative to the continuum. A figure showing the wavelength--dependent r.m.s. is presented in Appendix\,\ref{sec:1033correction} where relevant  pipeline artefacts are also detailed. 

\begin{figure}
    \centering
    \includegraphics[angle=0,width=\columnwidth]{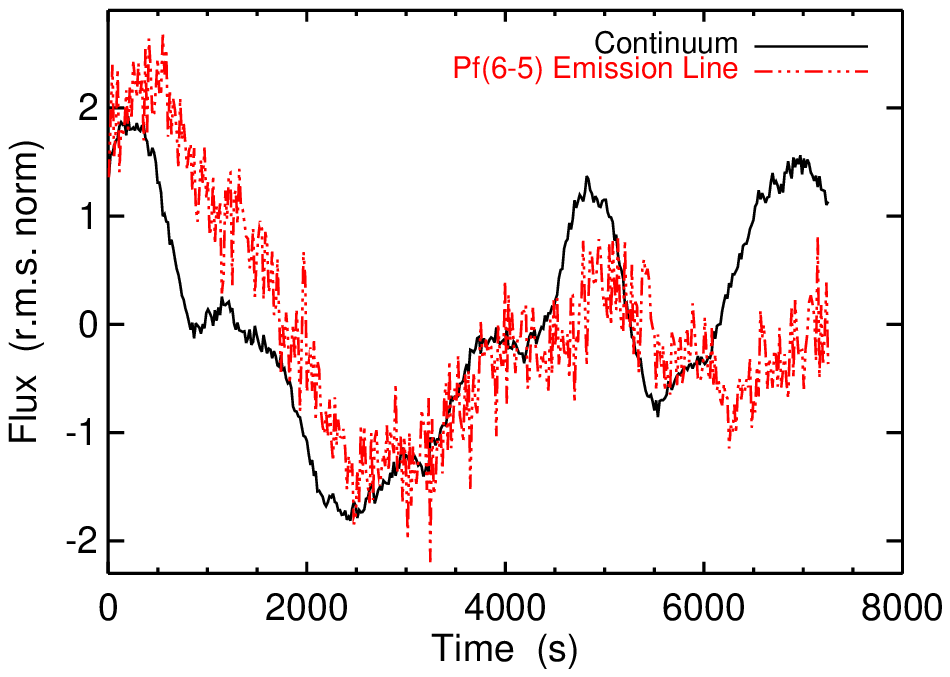}
    \caption{H\,Pf\,(6--5) line (7.35-7.59\,\micron) and adjacent continuum (7--7.25\,\micron) light curves, binned by a factor of 10, mean-subtracted and scaled by the respective scatter in each for cross-comparison.}
    \label{fig:cont_em_lc}
\end{figure}

\subsubsection{Timing: An Emission Line Time Lag}
\label{sec:timinglagresults}

We next searched for time lags in various spectral features. Light curves in the emission lines were found to show a similar variability pattern to the white-light data. Fig.\,\ref{fig:cont_em_lc} illustrates this with (i) a continuum-subtracted light curve aggregated over the strongest emission line Pf(6-5)\,$\lambda$7.35--7.59\,\micron\ and (ii) the adjoining line-free continuum region (7--7.25\,\micron). These have lower S/N than the white-light curve in Fig.\,\ref{fig:cont_em_lc}, as expected given their narrow wavelength span, but match the white-light profile well otherwise. 

Evident to the eye is an indication of a possible delay in the emission line relative to the continuum. In order to quantify this, we undertook a detailed timing analysis of the two light curves.

\begin{figure}
    \centering
    \includegraphics[width=0.5\textwidth]{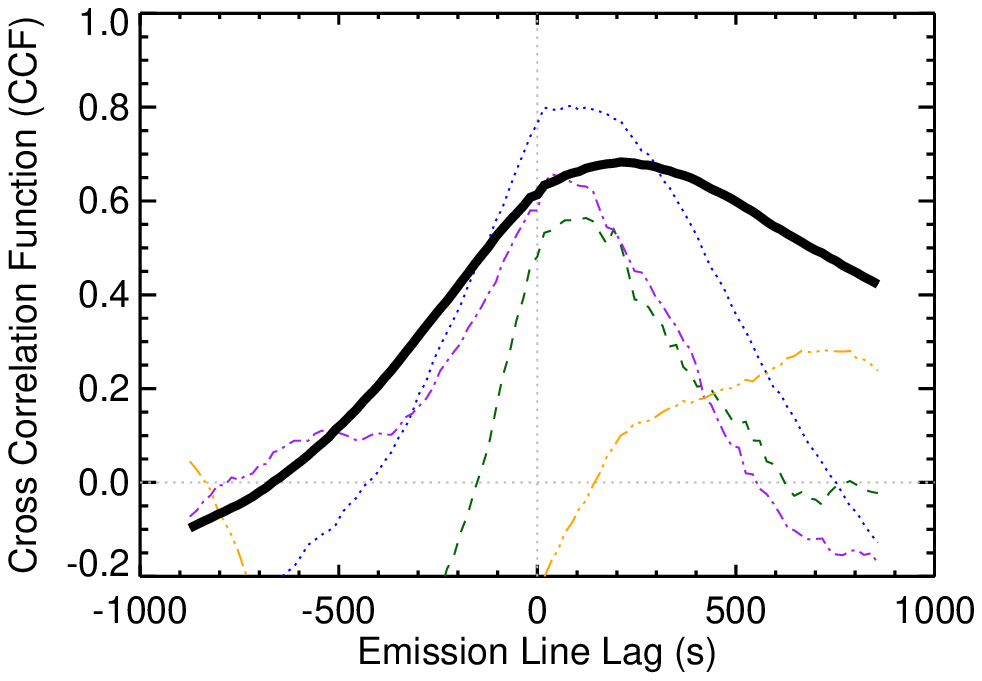}
    \caption{Cross correlations between the continuum and the Pf(6--5) line. The thick black line refers to the full light curve, and the four thin broken curves denote the CCFs for four consecutive and independent time segments each approximately 1,800\,s long. A positive value of the lag on the x-axis implies the Pf line lagging the continuum.}
    \label{fig:ccfs}
\end{figure}

We began by computing cross-correlation functions (CCFs), which quantify the strength of linear correlation between two time-series data sets, allowing for a relative lag \citep[cf. ][]{Gandhi-2010}. Fig.\,\ref{fig:ccfs} shows CCFs between the 7\,\micron\ continuum and the strongest Pf(6--5) line. The results are shown using the full light curves, as well as shorter independent time segments. In general the correlations are strong, and consistently show a central peak lying at positive emission-line lags (with an associated delay $\approx$\,50--300\,s following the continuum), though the broad width of the peak makes it difficult to constrain the lag well. One time segment also shows a peak different from the others (the orange curve in the figure with the weakest CCF strength). This refers to the final $\approx$\,30\,min of data, when the emission line response weakens substantially (cf. Fig.\,\ref{fig:cont_em_lc}); we will return to this weakening of the response shortly.

In order to model and constrain the lag better, we next attempted transfer function modelling. Many works have demonstrated the effectiveness of such an approach for timing analyses of active galactic nuclei (AGN) as well as XRBs \citep[e.g. ][]{peterson93, hynes98, zu11}. This method takes a driving input light curve $\mathcal{I}(t)$ and convolves it through a transfer function $\Phi(t)$ to predict the smeared and lagged output time-series $\mathcal{O}(t)$:

\begin{equation}
  \mathcal{O}(t) = N \int{\mathcal{I}(t')\, \Phi(t-t') \,dt'}
  \label{eq:convolution}
\end{equation}

\noindent
Here, $N$ is a normalising factor, which is a constant for stationary time series. For our purposes, the continuum is the driver input light curve $\mathcal{I}(t)$, and the model output $\mathcal{O}(t)$ is an estimator of the observed emission line light curve. 

We present results under the assumption of a Gaussian transfer function, which allows for a characteristic lag ($\tau$) together with a smearing length ($\sigma$) of the continuum light curve as it is reprocessed in some extended medium. As we will discuss later, our data do not allow a clean physical interpretation of the transfer function parameters. But for the purposes of aiding the reader, one may imagine $\tau$ as representing a mean time lag due to light-travel or other ionic physical processes within a gaseous shell surrounding the source, and $\sigma$ representing the physical extent of such a shell:  

\begin{equation}
  \Phi(t')=\frac{1}{\sqrt{2\pi}\sigma}e^{-\left( \frac{t'-\tau}{2\sigma} \right)^2}
  \label{eq:transfer}
\end{equation}

\noindent
Numerically, the transfer function is represented by a kernel of finite length, and we found consistent results for kernels ranging between 500--1,000 time bins. 

Parameter estimation was carried out through a maximum likelihood (ML) technique. This computes the log-likelihood that the observed time-series fluxes can be jointly reproduced by the transfer function model. In practice, the negative log-likelihood is minimised, and then sampled through a Markov Chain Monte Carlo (MCMC) sampling code, for which we utilised {\sc emcee} \citep{emcee} with 32 walkers, 5,000 samples, and a burn-in period of 200. Constraints on the model parameters are implemented through the use of uniform prior probabilities which modulate the likelihood functions. Specifically, negative values of $\tau$ and $\sigma$ were disallowed.\footnote{we disallowed $\sigma$\,$\le$\,1\,s to ensure that the transfer function remains numerically tractable over all reasonable parameter ranges.} 

We initially found that the above model did not produce an acceptable fit because of the aforementioned diminishing emission-line response with time, becoming substantially weaker toward the end of the observation (cf. Fig.\,\ref{fig:cont_em_lc}). This could be accounted for with a simple modification,  making the normalisation factor $N$ time-variable, $N\equiv N(t)$. An exponentially diminishing dependence was found to be sufficient to first order: $N(t)=e^{-t/\tau_{\rm abs}}$. The need for this parameter implies suggests that the time-series response is non-trivial, either non-linear or non-stationary, and we discuss possible interpretations of $\tau_{\rm abs}$ below.

\begin{figure}
    \centering
    \includegraphics[width=\columnwidth]{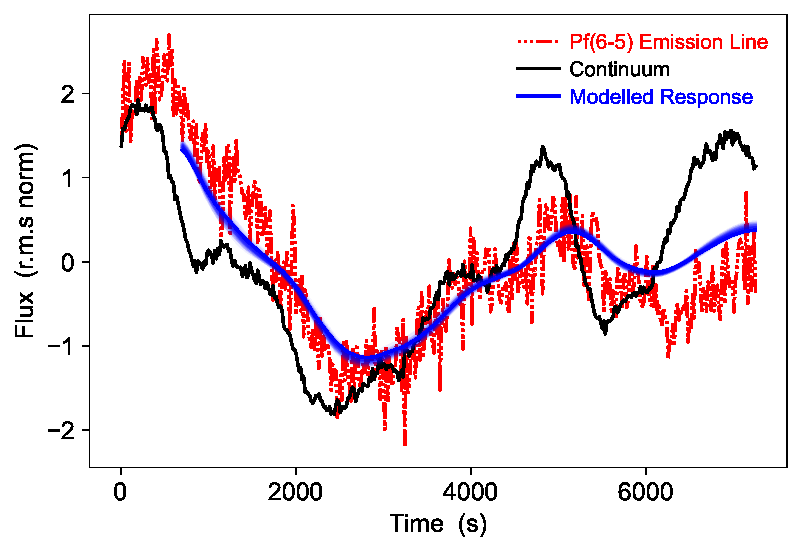}
    \includegraphics[width=\columnwidth]{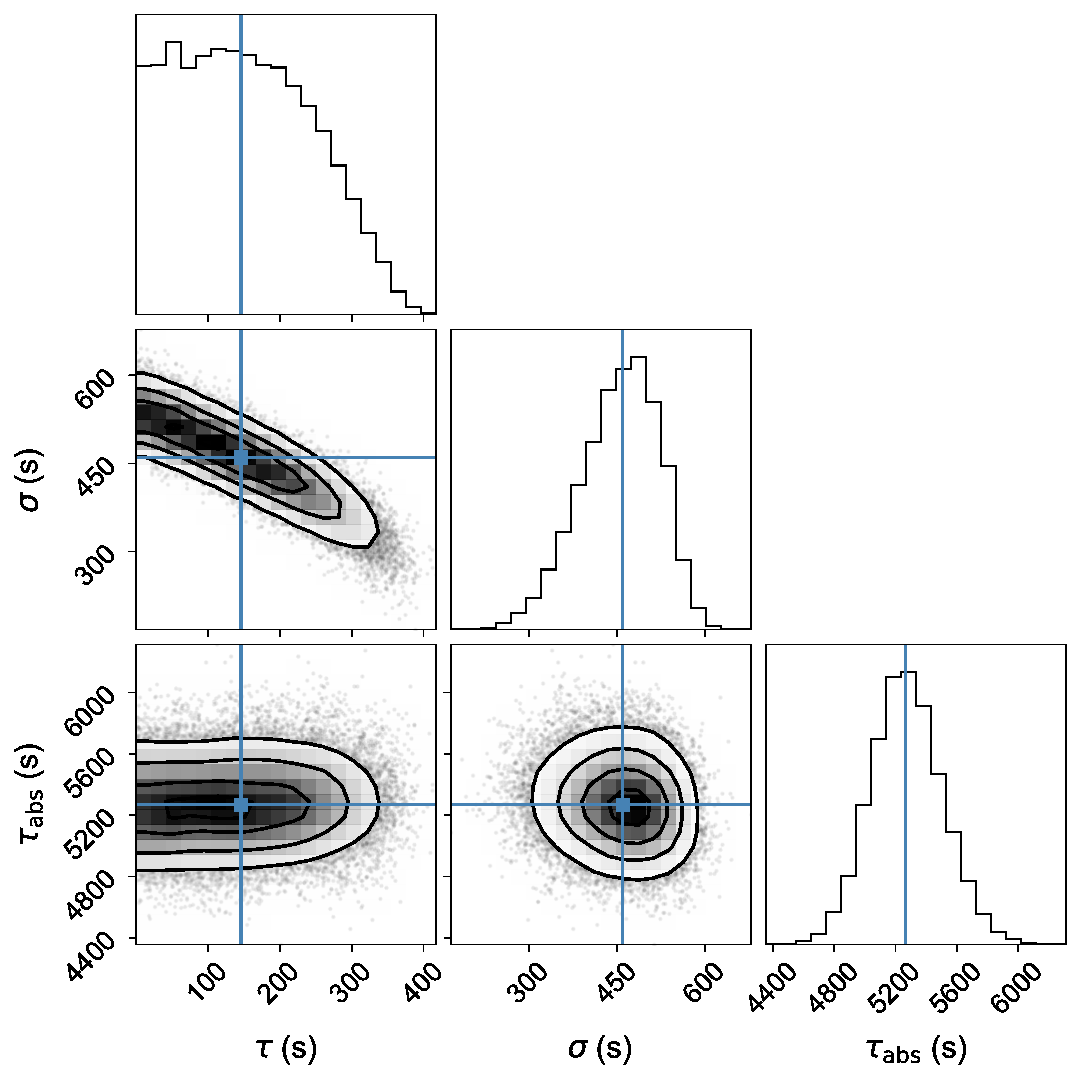}
    \caption{Results from the MCMC lag fitting. (Top) Emission line light curve (red), continuum light curve (black), and the smeared and delayed response of the continuum curve (blue; 100 randomly chosen best-fit samples). (Bottom) Corner plot, showing the distribution of fitted convolution parameters $\tau$ (mean lag), $\sigma$ (smearing width) and the factor characterising the timescale over which the response becomes non-stationary ($\tau_{\rm abs}$). 
    }
    \label{fig:lagfit}
\end{figure}

The results are shown in Fig.\,\ref{fig:lagfit} which plots a binned version of the respective light curves for display purposes, overlaid with several MC samples around the best modelled response. The model reproduces the overall shape of the emission line light curve well beyond $\approx$\,700\,s. Earlier times cannot be modelled because of the finite length of the convolution kernel, which means that reproducing the initial portion of the light curve would require knowledge of the driving input continuum before the start of our observation. 

The mean lag $\tau$ is found to be 1.5\,(\p\,0.1)\,$\times$\,10$^2$\,s, 
with the quoted uncertainties being the 16$^{\rm th}$ and 84$^{\rm th}$ percentiles of the marginalised sample. There is a strong degeneracy between $\tau$ and $\sigma$, the best fit for the latter centering around 4.5$_{-0.8}^{+0.6}$\,$\times$\,10$^2$\,s. 
This degeneracy is a result of the short duration of our observation probing only a few distinct features and peak-to-peak cycles in the light curve. In other words, there are insufficient handles in our sampled dataset to distinguish between lagging and smearing effects which can qualitatively mimic each other. 

The final free parameter $\tau_{\rm abs}$ is fitted to be 5.2\,(\p\,0.2)\,$\times$\,10$^3$\,s. In the light curves shown in Fig.\,\ref{fig:lagfit}, it is apparent that the magnitude of the emission line flux variations diminishes on such a timescale, relative to those seen in the continuum. This parameter still fails to perfectly replicate the variations on the longest timescales, with the models overpredicting the data beyond $\approx$\,5,500\,s, though the overall shape and lags appear to be well reproduced. Interpretations of $\tau_{\rm abs}$ include a change in the emission line response caused by patchy and changing absorption (as was also inferred in the XRB V404\,Cyg; \citealt{motta17, walton17}). Specifically, if obscuration between the nucleus and the emission-line region increases over the course of the MIRI observation, this could manifest as an increasing optical depth that we parameterise as $\tau_{\rm abs}$. Alternatively, there could have been a systemic change in the source variability at late times, with the lag itself lengthening in a non-stationary manner toward the end of our observation. Finally, we cannot rule out the more mundane possibility that $\tau_{\rm abs}$ may simply be a `fudge' factor required by our current limited sampling of the observed time-series. 

Most other emission lines did not reveal any significant lag due to their comparative faintness. The only other emission line to show weak evidence for a lag was the Hp\,(7--6) line, which should dominate the next strongest observed line feature near 12.4\,\micron. In this case, we found $\tau$\,=\,1.0$_{-0.7}^{+0.9}$\,$\times$\,10$^2$\,s, $\sigma$\,=\,2.1$_{-0.8}^{+0.8}$\,$\times$\,10$^2$\,s, and $\tau_{\rm abs}$\,= 
6.5$^{+0.5}_{-0.6}$\,$\times$\,10$^3$\,s. 
This is qualitatively consistent with the results from the Pfund line (including the degeneracy between $\tau$ and $\sigma$), but with a shorter smearing length by a factor of about 2. The increased r.m.s. in the red region of the spectrum (\S\,\ref{sec:obs}) almost certainly impacts accurate estimation of uncertainties for this line. 

\section{Discussion}
\label{sec:discussion}

\subsection{Mid-Infrared Luminosity}
\label{sec:powerdiscussion}

The observed MIRI continuum is at a flux density  $F_\nu$\,$\approx$\,70\,mJy at 8\,\micron, corresponding to a monochromatic MIR luminosity $\lambda L_{\lambda\,(8\,\mu m)}$\,=\,2.8\,(\p 0.7)\,$\times$\,10$^{35}$\,erg\,s$^{-1}$, or an integrated luminosity over the full MIRI spectral range of $L_{4-14}$\,=\,4.1\,$\times$\,10$^{35}$\,erg\,s$^{-1}$. 

The optical extinction to \grs\ has been estimated by \citet{chapuis04} based upon CO observations of interstellar molecular clouds along the line-of-sight and assuming standard conversions between CO and H column densities observations \citep{strongmattox96, rodriguez95}, together with known correlations between interstellar visual extinction and X-ray dust scattering optical depth \citep{predehlschmitt95}. The estimated interstellar extinction is \av\,=\,19.6\,\p\,1.7\,mag which can be used to deredden the observed spectrum, if an extinction curve spanning the optical to MIR wavelength range is known. To achieve this, we follow \cite{rahoui10}, who combined a standard selective extinction $R_{\rm V}$\,=\,3.1\,mag value with the optical--NIR \citep{fitzpatrick99} and NIR--MIR \citep{chiartielens06} extinction curves. The resultant $A_{\rm \lambda}/A_{\rm V}$ values covering the MIRI spectral range are listed in Appendix\,\ref{sec:extinction}. For reference, $A_8$\,=\,0.8\,mag at 8\,\micron, and MIR extinction peaks around $A_{10}$\,$\approx$\,2.1\,mag within the silicate absorption feature. 

Applying these extinction corrections results in an integrated intrinsic luminosity $L_{4-14}^{\rm int}$\,=\,1.1\,$\times$\,10$^{36}$\,erg\,s$^{-1}$. The Eddington luminosity of the source for radiation impinging on ionised gas is $L_{\rm Edd}$\,=\,1.4\,$\times$\,10$^{39}$\,erg\,s$^{-1}$, using $M_{\rm BH}$\,=\,11.2\,\Msun\,(Table\,\ref{tab:pars}). So a significant fraction $L_{4-14}^{\rm int}$/$L_{\rm Edd}$\,=\,$7\times10^{-4}$ of the Eddington power is being radiated in the MIRI spectral range during our observations. 

\begin{figure*}
    \centering
    \includegraphics[angle=0,width=\textwidth]{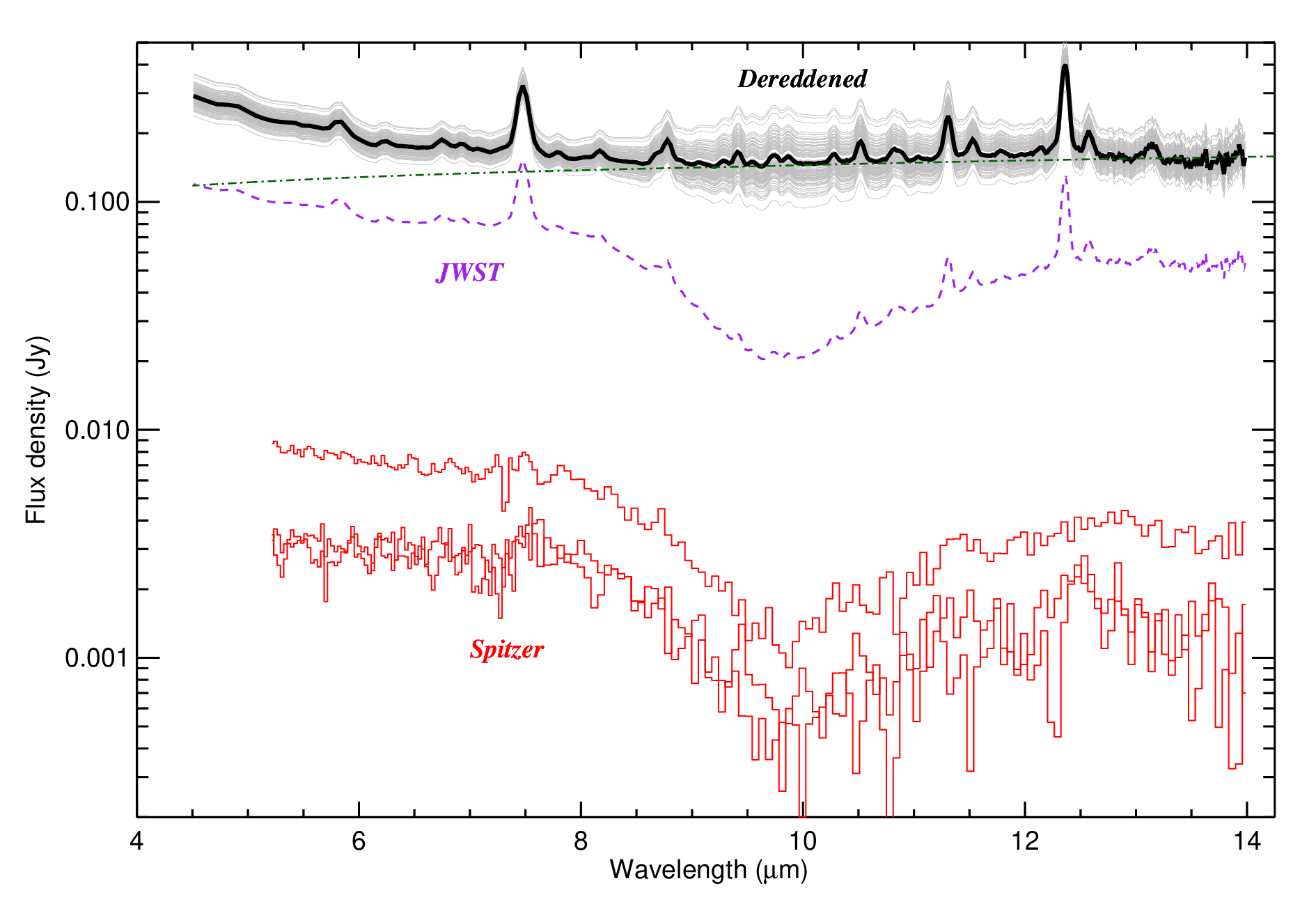}
    \caption{The observed MIRI/LRS spectrum of \grs\ in dashed purple, shown together with dereddened spectra (top), and three archival \spitzer/IRS observed spectra (bottom red) from MJD 53851.4117545, 53511.6928638 and 53299.219684. The spectra in grey have been dereddened assuming interstellar (ISM) \av\ values sampled from a normal distribution with mean 19.6\,mag and dispersion 1.7\,mag. The thick black curve represents the mean intrinsic spectrum corrected for ISM extinction. The green dot-dashed curve represents an example thermal bremsstrahlung scenario for the continuum (\S\,\ref{sec:bremss}).}
    \label{fig:spitzer}
\end{figure*}

Correcting the entire MIRI range for extinction with the above prescription will effectively correct the curvature arising from silicate absorption. This is shown in Fig.\,\ref{fig:spitzer}, where several dereddened spectra are plotted together with an average dereddened spectrum (assuming a range of extinction values, sampled assuming a normal distribution using the known \av\ uncertainty). We discuss the shape of this intrinsic spectrum below. 

Previous MIR spectral observations of the source have included several by the \spitzer\ mission \citep{spitzer}, with the source being detected by the IRS instrument \citep{spitzerirs} during 8 observations spanning the years 2004--2006. Their detailed analysis has been published by \citet{rahoui10}, and three representative epochs are plotted in Fig.\,\ref{fig:spitzer}. These were downloaded from the \spitzer\ IRS Enhanced Products archive, which provides background-subtracted, flux-calibrated and merged spectra for each observation epoch.\footnote{\url{https://irsa.ipac.caltech.edu/data/SPITZER/docs/irs/irsinstrumenthandbook/84}}

The continuum profiles of all the IRS spectra approximately match the MIRI data, albeit with much lower S/N. The median observed IRS flux density was 4\,mJy with all the historical flux levels being nearly identical, except for one epoch on MJD 53851.41175 when the source was found to be closer to 7 mJy (this is the brightest \spitzer\ epoch shown in Fig.\,\ref{fig:spitzer}). Going further back, the mean flux of the source when it was observed photometrically in 1996--1997 with the {\em ISO} mission was $\sim$\,5--10\,mJy \citep{fuchs03}, very similar to the \spitzer\ era. By contrast, the median flux observed with \jwst\ in 2023 is close to 50\,mJy, a factor of $\approx$\,5--13\,$\times$ brighter than historical values. We will discuss the implications of these long-term changes in \S\,\ref{sec:finaldiscussion}.

\subsection{Narrow Emission Lines}
\label{sec:linediscussion}

\begin{figure*}
    \centering
    \includegraphics[angle=0,width=\textwidth]{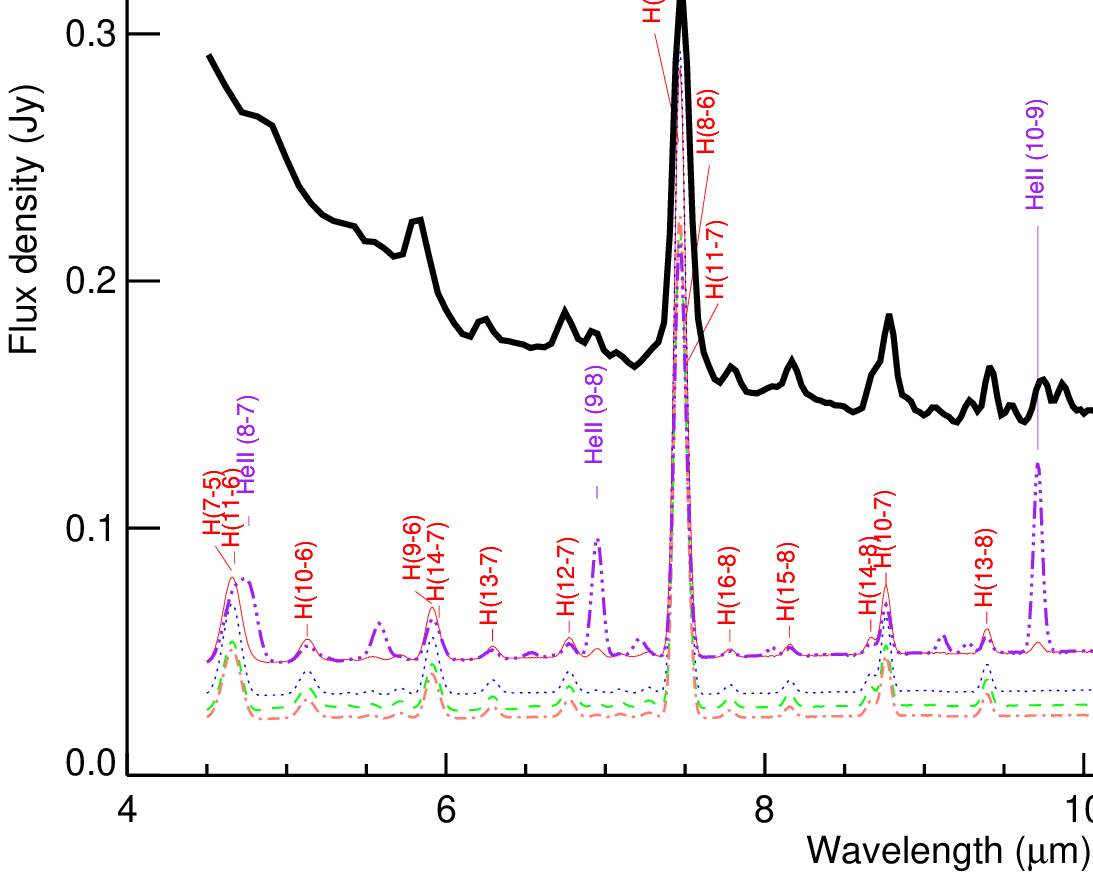}
    \caption{MIRI/LRS dereddened spectrum of \grs\ compared to several {\tt NEBULAR} model spectra with annotated temperatures (in K) and densities (cm$^{-3}$).}
    \label{fig:spec_with_nebular}
\end{figure*}

In Fig.\,\ref{fig:spec_with_nebular}, several model spectra are overplotted together with our median dereddened source spectrum. The models are recombination line spectra for H and He, for a range of annotated temperatures ($T$\,=\,10,000--20,000\,K) and densities ($n_{\rm H}$\,=\,10$^{3-8}$\,cm$^{-3}$). The spectra were produced with the {\tt NEBULAR} code \citep{nebular}, assuming Case B recombination and 10\%\ He abundance by part. The models have been convolved with a Gaussian kernel of spectral resolution, changing with wavelength according to \citet{kendrew15} in order to approximate the unresolved line widths as seen by MIRI. 

The most immediate result apparent to the eye is that many of the emission lines, especially the strongest ones, are consistent with simple high-order H recombination transitions of the Pfund ($n_l$\,=\,5), Humphreys ($n_l$\,=\,6) and higher series, up to $n_l$\,=\,8 at least. Whilst it is clear even to the eye that the relative model line ratios of many of the strongest lines approximate those seen in the data, it is worth emphasising again that the low spectral resolution results in overlapping line associations, thus precluding robust identification in many cases. So all of our line identifications and inferred flux measurements of individual features should be treated with some caution. A follow-up MIRI medium-resolution spectrum (MRS mode) should be able to resolve many of these degenerate identifications. 

Several emission features with no current identification are also annotated therein. Some of these have potential matches with \hei\ transitions found in the linelist of \citet{cloudy23}, but typically only at very high excitation, so we have not included them herein. {\tt NEBULAR} does not include \hei\ transitions in the MIRI wavelength range. 

The location and nature of the line-forming region is currently unknown, but the model spectra allow a constraint on temperature under the assumption of Case\,B recombination. Specifically, high temperatures $T$\,$\gtsim$\,20,000\,K in the MIR line emission region are excluded by the absence of prominent \heii\ emission lines; cf. the strong prediction in the case of \heii\,(10--9) feature at 9.7\,\micron\ which we do not see. There is a weak nearby feature present in the data ($\lambda$\,=\,9.735\,\micron) which, if associated with \heii\,(10--9), together with a similar possible identification of \heii\,(9--8)\,$\lambda$\,6.9\,\micron\ would be consistent with a temperature of $T$\,$\approx$\,17,000\,K or lower. Though these recombination models provide surprisingly good qualitative descriptions of the data, it should be kept in mind that the XRB environment is likely more complex, and subject to the effects of a strongly accreting central engine. We touch upon more realistic photoionisation simulations in \S,\ref{sec:lboldiscussion}.

\subsection{Dust Features}
\label{sec:dustdiscussion}

Spectral line indicators for the presence of dust include broad emission bands, 
such as 
Polycyclic Aromatic Hydrocarbons (PAHs) which comprise large carbonaceous molecules and are commonly found in the ISM, in star-forming galaxies and around AGN \citep[e.g., ][]{allamandola89, esquej14}. Vibrational modes in these molecules can be excited by ultraviolet radiation, which then results in infrared emission. There are several bands expected in the MIRI range -- around 6.2, 7.7 and 11.3\,\micron, in particular. The centroid wavelengths and profile shapes of these features can vary depending upon the size and ionisation state of the molecules present. 

Detection of these features in the MIRI spectrum is complicated by the low spectral resolution of our data. Around all three wavelengths, there are H recombination lines -- H(13--7), H(16--8) and H(9--7), respectively, that could cause confusion. The emission feature at 11.3\,\micron\ is the strongest of these, and its flux ratio relative to the other lines suggests that it is dominated by H(9--7)\,$\lambda$11.303\,\micron\ recombination. PAH features are also thought to be destroyed in bright X-ray dominated regions \citep{voit92}, and as we discuss later, \grs\ probably remains in an intrinsically active and luminous state. 

For all the reasons above, we are cautious of claiming detection of PAH features in the MIRI spectrum. Their detection was claimed by \citet{rahoui10} in the IRS data, where the recombination lines were also weaker. Their report of PAH features during the \spitzer\ epochs is thus more reliable. The \spitzer\ spectra have been further analysed by \citet{harrison14}, who confirmed the presence of PAHs, though they highlighted that these features need not necessarily originate in close proximity to the source. 

Analysing the fluxes of the putative PAH features in the MIRI spectra, we find that the flux of the emission line at 7.8\,\micron\ (which could potentially be PAH\,7.7 or could be H\,(16--8) instead) is very similar to that reported by \citet{rahoui10}. If this really is attributable to PAHs, then it would require a steady PAH flux since $\sim$2004, despite the dramatic change in continuum flux. However, we also highlight the fact that all the strongest MIRI emissions lines and any lines near expected PAH wavelengths are narrow (FWHM\,$<$\,0.1\,\micron; cf. Table\,\ref{tab:emlines}). By contrast, PAHs are generally measured to be significantly broader (FWHM\,$>$0.1--0.2\,\micron) and often with skewed profiles, even in low S/N datasets (e.g., \citealt{allamandola89, rahoui10, armus23}), arguing that all of the relevant MIRI features are instead better explained as ionic (gas) recombination lines.

We thus rule out any substantial {\rm brightening} of PAHs since the \spitzer\ observations in 2004--2006, with our data suggesting that PAHs may have since even been {\em destroyed}, presumably due to a change in the ionising conditions of the PAH emission region. If so, not only does this place the site of the PAHs to be the vicinity of \grs, it also sets a conservative upper-limit of $\approx$\,17\,years for the dust destruction timescale. An MRS observation will be able to provide a more sensitive test of this hypothesis by helping to cleanly deconvolve the recombination lines.

Additional weak and broad unidentified infrared bands are often taken to be signatures of the presence of dust \citep{chiartielens06}. Our dereddened spectrum is not perfectly flat (Fig.\,\ref{fig:spec_with_nebular}), and we cannot rule out the presence of such broad features (e.g. around 12\,\micron). 

\subsection{Timing}
\label{sec:timingdiscussion}

\jwst's high sensitivity has enabled the detection of low amplitude, but significant, flux variations of order 1\,\%\ (Fig.\,\ref{fig:lc}). Any characteristic variations in the light curves are on long timescales ($\gtsim$\,1000\,s, at least), and the corresponding PSD rises steeply to low frequencies. Steep red-noise PSDs with slopes approaching $\beta$\,=\,--2 are rare in XRBs, but have been observed in a few other sources, e.g. during the bright 2015 outburst of V404\,Cyg in the optical \citep{gandhi16}. Such long timescales exceed the light-travel time across the systems in both cases, and appear more akin to accretion-induced variations. Both V404\,Cyg and \grs\ were probably also observed at relatively high $L/L_{\rm Edd}$, though variability and obscuration render bolometric luminosity measurements difficult (cf.\,\S\,\ref{sec:lboldiscussion}). 

The emission line lag that we find is also exceptionally long. Studies of reprocessing in XRBs have focused on the inner accretion zones -- either the accretion disc or inner jet \citep[e.g., ][]{Hynes-2003, Gandhi-2008, Gandhi-2010, Casella-2010, paice19, vincentelli19, tetarenko21}, or on the companion star \citep{obrien02, munozdarias07}, where characteristic lags are of order seconds. Short near-infrared (NIR) continuum lags with respect to X-rays have also been found in \grs\ \citep{Lasso-Cabrera-2013}. Common patterns of X-ray and NIR rise and decoupling of flares on timescales of $\sim$\,30\,min were found in the early days of multiwavelength follow-up of the source, and ascribed to synchrotron plasma ejecta \citep{Mirabel-1998, eikenberry98_diskjet}. With regard to the emission lines specifically, continuum and line flux correlations were identified in the NIR by \citet{eikenberry98_irflares} and interpreted as evidence for radiative line pumping, though no lags could be identified in those data. 

Our direct measurement of an emission line lag is the first such evidence of MIR line flux `reverberation' in an XRB, to our knowledge. The mean lag that we measure is $\tau$\,$\approx$\,150\,s, but there is strong degeneracy with the smearing timescale kernel which is a factor of $\approx$\,3 longer. Moreover, the transfer function itself appears to be non-stationary in time (\S\,\ref{sec:timingresults}). All of this impacts our ability to pin down the exact lag in our data set, though it is nevertheless consistent with being several hundred seconds. A modestly longer observation probing a few additional flux rise and fall cycles should be able to determine the lag robustly. 

The line--continuum correlation that we observe does not necessarily imply a causal relationship. Instead, it is likely that both the MIR continuum and lines are driven by higher energy (UV) and X-ray radiation from the central source. If one wishes to determine the true lag of the emission lines with respect to the central high energy radiation, an {\em additional} lag related to the light-travel time from the nucleus 
will need to be included. If the MIR originates near the outer accretion disc extending close to the Roche Lobe of the secondary star, this additional lag will be of order 200\,s (cf. Table\,\ref{tab:pars}). 

Furthermore, any finite recombination times will also introduce time delays, further complicating the interpretation of the lags. In order to evaluate  this, we need a constraint on the gas density, which we turn to next. 

\subsection{The Gas Density}
\label{sec:densitydiscussion}

Constraining the gas density $n_{\rm H}$ is key to determining the nature of the source of MIR emission. Here, we consider four complementary estimates on $n_{\rm H}$ and what we can learn from these independent considerations. It is worth bearing in mind that all are subject to the caveat that we are likely only probing the optically-thin portion of the emitting gas, and that the geometry of the medium may well be non-uniform (i.e. clumpy and with a non-uniform sky-covering factor). 

\subsubsection{The MIR Line-Emitting Region:\\\hspace*{0.8cm}Constraints from Line Fluxes}
\label{sec:linefluxconstraints}

The gas density influences line emissivities, 
so a constraint on \nh\ can be obtained from the observed line fluxes assuming ionisation equilibrium within some radius $r$. Standard recombination theory connects the rate of Lyman continuum photons, \nlyc, i.e. photons with energy $E$\,$>$\,13.6\,eV, to the fluxes of various elemental transitions. A commonly adopted relation between the Balmer line H$\alpha$\,$\lambda$6563\,\AA\ and \nlyc\ is

\begin{equation}
    L({\rm H\alpha}) = 1.36 \times 10^{-12}\,\times\,Q({\rm H^0})
\end{equation}

\noindent
where $L({\rm H\alpha})$ is expressed in erg\,s$^{-1}$ and \nlyc\ in photons\,s$^{-1}$ \cite[cf., ][]{bouwens16, kennicutt98}. The above relation has only a weak dependence on temperature and abundance. 

The dereddened flux of the strong H(6--5) Pfund $\alpha$ emission line in MIRI is $F_{\rm H(6-5)}$\,=\,1.1\,$\times$\,10$^{-12}$\,erg\,s$^{-1}$\,cm$^{-2}$. Uncertainties on this flux are dominated by systematics related to potential confusion with weaker lines and disentangling of the continuum. By experimenting with varying underlying continua and a variety of modelled spectra with {\tt NEBULAR}, we estimate that these could contribute as much as 25\,\%\ uncertainty.

The line luminosity in \ha\ can next be predicted from recombination theory: the typical Case\,B Pfund--to--Balmer line ratio is \pfa:\ha\,$\approx$\,0.0246 \citep{nebular, cloudy23, hummerstorey87} which again is not particularly sensitive to the physical conditions over the range inferred above. Together with our systematic uncertainty estimate, we predict $L$(\ha)\,$\approx$\,4.7\,(\p\,1.2)\,$\times$\,10$^{35}$\,erg\,s$^{-1}$ which, in turn, yields log(\nlyc)\,=\,47.5\,\p\,0.1. 

Finally, assuming complete ionisation equilibrium within some radius $r$ yields the Lyman continuum photon rate as

\begin{equation}
    Q(H^0) = \frac{4}{3}\pi r^3 n_e n_{\rm H} \alpha_{\rm B}
\end{equation}

\noindent
where $\alpha_{\rm B}$ is the recombination coefficient and the electron density $n_e$ is taken to be identical 
to \nh. For the inferred temperature constraint (\S\,\ref{sec:linediscussion}), we expect $\alpha_{\rm B}$\,$\approx$\,2.5\,$\times$\,10$^{-13}$\,cm$^3$\,s$^{-1}$ \citep{osterbrock}. 

A reasonable ansatz for the radius $r$ is the size of the outer accretion disc. At $\approx$\,200\,lt-sec, this is of similar order to our observed time lag (\S\,\ref{sec:timingdiscussion}). Adopting this value of $r$, we finally obtain  \nh\,$\approx$\,4\,$\times$\,10$^{10}$\,cm$^{-3}$ in the MIR emission line zone.

\subsubsection{The MIR Line-Emitting Region:\\\hspace*{0.8cm}Constraints from Flux Ratios}

The relative line strengths, in particular the comparatively weak strength of \heii, have constrained the mean temperature of the emission line region to be $T$\,$<$\,20,000\,K (\S\,\ref{sec:linediscussion}). 

By contrast, the models in Fig.\,\ref{fig:spec_with_nebular} are not strongly sensitive to space density, with little distinction between spectra up to the maximum densities that can be simulated in {\tt NEBULAR}, $n_{\rm H}$\,=\,10$^8$\,cm$^{-3}$. In denser media with $n_{\rm H}$\,$\gtsim$\,10$^{10}$\,cm$^{-3}$, the flux ratios of higher-to-lower order transitions of H can change significantly \citep[e.g., the H(9--7):H(7--6) ratio increases by factors of several to exceed unity; ][]{kwanfischer11, franceschi24}, which we do not observe. This places a constraint of $n_{\rm H}$\,$\ltsim$\,10$^{10}$\,cm$^{-3}$ for the emission line zone. 

This gas density, based upon the observed line {\em ratios}, is of the same order-of-magnitude as the \nh\ value inferred from the line {\em fluxes} in the preceding section, though there is mild disagreement between the two. However, it should be noted that values of \nh\,$\gtsim$\,10$^{10}$\,cm$^{-3}$ or so are relatively high, so the assumption of pure Case B ionisation equilibrium that we have relied upon so far may not be valid, with collisional and optical depth effects needing to be accounted for. We will return to this point in \S\,\ref{sec:lboldiscussion}.

\subsubsection{The MIR Continuum-Emitting Region:\\\hspace*{0.8cm}A Thermal Bremsstrahlung Origin?}
\label{sec:bremss}

The origin of the MIRI continuum is attributed to physical regions at the system outskirts in several plausible scenarios -- either near the outer accretion disc or at the companion star \citep{rahoui10}. Our r.m.s. analysis (Fig.\,\ref{fig:lc}) also shows a drop in variance at the positions of the emission lines, as expected if the line emitting gas is not cospatial with (and likely more extended than) the continuum. So it is instructive to derive estimates of \nh\ for the continuum emitting gas to compare with the lines. 

Here, we focus on the specific scenario of thermal bremsstrahlung radiation from the wind. The wind has been proposed as the obscuring medium in the current obscured state of the source \citep{miller20}. Constraints on \nh\ are possible from the MIRI data alone in such a scenario, but we stress that such an origin cannot be definitely proven in our data, though the discussion below will show why it is plausible. Other possible interpretations will be outlined in the following sections. 
The observed thermal bremsstrahlung flux density can be predicted as \citep{kellogg75_bremss}: 

\begin{eqnarray}
    F_\nu = A\,\sqrt{T}\, g(\nu,T)\,e^{-\frac{h\nu}{kT}}
    \label{eq:bremss}
\end{eqnarray}

\noindent 
Here, $g(\nu, T)$ is the Gaunt factor of order 1 and $A$ is a normalising factor,  

\begin{equation}
    A=\frac{2.0\times 10^{-18}\int{n^2 dV}}{4\pi d^2}
    \label{eq:bremssnorm}
\end{equation}

\noindent
with all lengths expressed in units of cm, and the constant factor appropriate for flux density in Jy. Here we utilise the implementation of the bremsstrahlung model in the {\tt xspec} code \citep{xspec}, with Gaunt factors from the polynomial fits of \citet{bremss_xspec_gaunt}. 

The wind must be hot, if launched by the accretion disc, but our data do not provide any constraint on the temperature of the continuum-emitting region, except for the fact that the flat dereddened MIRI SED implies the bremsstrahlung spectral cutoff must lie at frequencies higher than the MIR. In such a case, the emission measure ($\int{n^2 dV}$) is the main parameter governing the normalisation of the bremsstrahlung model to the MIRI spectrum. It is then possible to obtain an estimate of the mean wind density, once a characteristic radius is  assumed. 

\citet{rahoui10} used a Compton temperature of $T$\,=\,5.8\,$\times$\,10$^6$\,K. But even much lower values characteristic of the electron temperature in the emission line region ($T$\,$\sim$\,10$^4$\,K) change the inferences below by only a factor of $\sim$\,2. Fitting a bremsstrahlung model to our dereddened MIRI spectrum at 10\,\micron, and assuming a mean spherical volume $V$\,=\,4/3\,$\pi$\,$r^3$, with $r$ representing the outer disc (200\,lt-sec), we find $n_{\rm H}$\,=\,(1--3)\,$\times$\,10$^{11}$\,cm$^{-3}$. One such model is plotted in Fig.\,\ref{fig:spitzer}.

\subsubsection{A cross-check from the X-ray Absorption Column}

Finally, a simple consistency cross-check comes from the column density $N_{\rm H}$ estimated from X-ray observations, combined with our estimate of the extent of the obscuring medium:

\begin{equation}
    N_{\rm H} \approx n_{\rm H} \times r
    \label{eq:columndensity}
\end{equation}

We do not present X-ray data herein, but the source has clearly faded across the MAXI energy range up to $\approx$\,20\,keV, at least (Fig.\,\ref{fig:longterm}). Previous studies have found  Compton-thick columns in the hard state \citep[e.g., ][]{balakrishnan21}. In early 2023, a column at the {\em low} end of optical depths with $N_{\rm H}$\,=\,3\,(\p\,0.5)\,$\times$\,10$^{23}$\,cm$^{-2}$ was reported by \citet{miller23_atel}. 

Using $r$\,=\,200\,lt-sec and a column density of $N_{\rm H}$\,=\,10$^{24}$\,cm$^{-2}$ barely in the Compton-thick regime yields a conservative space density of $n_{\rm H}$\,$\approx$\,2\,$\times$\,10$^{11}$\,cm$^{-3}$. If the space density has a radial profile decreasing away from the nucleus, then $n_{\rm H}$ closer in to the nucleus will be higher. Thus, our estimate should be considered a conservative mean value for a uniformly distributed annular sphere of obscuring matter. 

\subsection{Summary constraints on gas density and mass}
\label{sec:mass}

The four separate estimates above yield constraints on the gas density ranging from \nh\,$\ltsim$\,10$^{10}$\,cm$^{-3}$ to 4\,$\times$\,10$^{10}$\,cm$^{-3}$\,in the emission line region, to (0.5--2)\,$\times$\,10$^{11}$\,cm$^{-3}$ inferred from the (MIR and X-ray) continua. These could all arise from the same medium (e.g., the wind), with the difference in densities arising from stratification within the medium. The lower density of the emission line region is consistent with the lines arising from the diffuse, outer fringes, while the denser, compact zones of the wind could be responsible for the obscuration and the MIR bremsstrahlung emission. 

All of these estimates imply that the mass $M$ of this gas surrounding the source must be high. For a uniform spherical distribution of gas: 

\begin{eqnarray}
    M \approx \frac{4}{3}\pi r^3 n_{\rm H} m_p f
\end{eqnarray}

\noindent
Taking a filling factor $f$\,=\,1 and $n_{\rm H}$\,=\,10$^{10}$\,cm$^{-3}$ (at the lower end of the above estimates appropriate for the emission line region), this yields $M$\,$\approx$\,2\,$\times$\,10$^{22}$\,kg\,=\,8\,$\times$\,$10^{-9}$\,\Msun. 

Alternatively, an outflowing annular shell is likely to be a better description of the geometry of the {\em wind} known to be present in this source. Assuming \nh\,=\,10$^{11}$\,cm$^{-3}$ for the bremsstrahlung-emitting wind and a terminal wind velocity $v_w$\,=\,10$^3$\,km\,s$^{-1}$ (cf.\,3,000\,km\,s$^{-1}$ was inferred from near-infrared spectroscopy carried out in May 2023 about 2 weeks prior to the \jwst\ observation;  \citealt{sanchezsierras23_atel}), we obtain an estimate of the mass outflow rate as

\begin{eqnarray}
    \dot{M} & \approx & 4\pi r^2 n_{\rm H} m_{\rm p} v_{\rm w} \Omega_{\rm w}\\
            & =       & 8\,\times\,10^{21}\,{\rm g}\,{\rm s}^{-1}\,\,=\,\,1\,\times\,10^{-4}\,{\rm M_\odot}\,{\rm yr}^{-1}.
\end{eqnarray}

\noindent
This is an extremely high mass-loss rate. Wind covering factors $\Omega_{\rm w}$\,$\ll$\,1 could moderate the above rates, but these require a fine-tuned geometry to maintain an obscured view of the source, as has persisted for $\approx$\,6\,years now since 2018. So $\Omega_{\rm w}$ is unlikely to be very small. A {\em toroidal} geometry should be dynamically more stable than a sky-covering spheroidal annulus, but such a configuration would still only change $\dot{M}$ by a factor of order unity. 

The strongest dependence in the mass-loss rate above is on radius ($\dot{M}$\,$\propto$\,$r^2$), for which we have assumed a plausible size corresponding to the outer disc but we cannot rule out smaller sizes. Reducing $r$ by a factor of 10 (say) would decrease $\dot{M}$ to $\sim$\,10$^{-6}$\,\Msun\,yr$^{-1}$ for a constant \nh. However, \nh\ should also scale down as $r^{-1}$ (from the X-ray column density estimator; Eq.\,\ref{eq:columndensity}) or as $r^{-1.5}$ (for a constant bremsstrahlung emission measure; Eq.\,\ref{eq:bremssnorm}). So the dependence of mass-outflow rate on radius is effectively weaker, scaling as $\dot{M}$\,$\propto$\,$r^{0.5-1}$.

Another inherent assumption is that the wind is unbound. If the MIR continuum probes a phase of the wind distinct from the fast-moving NIR component, $v_w$ could also be reduced, thereby moderating $\dot{M}$ substantially. We will revisit the impact of such assumptions in \S\,\ref{sec:finaldiscussion}.

\subsection{Other possible sources of the continuum}
\label{sec:othercontinuum}

The MIRI continuum cannot be attributed to the donor star, which should contribute no more than $\ltsim$\,1--10\,\% of the dereddened flux at any wavelength across the MIRI range (cf. \citealt{rahoui10}). The observed flat slope (Fig.\,\ref{fig:spitzer}) is also inconsistent with a simple blackbody. 

The outer disc of \grs\ is approximately 200\,lt-sec in radius (cf.\,Table\,\ref{tab:pars}). This is about (3--4)\,$\times$\,10$^6$ Gravitational radii for the black hole mass $M_{\rm BH}$\,=\,11.2\,\p\,1.7\,\Msun. Irradiation by nuclear X-rays and ultraviolet can heat the outer disc, flattening its SED relative to a blackbody \citep[e.g., ][]{diskir}. Efficient heating of a large outer disc may then partially account for the shape of the MIRI continuum. It would have to be a very stable outer disc in terms of radiated flux, with an r.m.s. of no more than a few per cent over $\approx$\,2\,hours (Fig.\,\ref{fig:cont_em_lc}). 

Another possibility is synchrotron radiation. Jetted emission has historically been very strong in this system, and studies of the radio activity during the obscured state have demonstrated continued, enhanced flaring \citep{motta21}. If the MIR continuum is dominated by a compact jet, the flat SED requires that the MIR lies in the optically-thick regime. This, in turn, constrains the optically-thick to -thin synchrotron spectral break (\nub) lying at frequencies higher than the MIRI range, which also places a constraint on the magnetic field ($B$) at the radiative base of the jet \citep{Blandford-1979}.

To estimate $B$, we follow the prescription in \citet{Gandhi-2011} for a homogeneous single-zone equipartition region at the base of the jet. Taking a lower limit on \nub\,$\ge$7.5\,$\times$\,10$^{13}$\,Hz ($\equiv$\,4\,\micron), a break flux density $F_{\nu_{\rm b}}$\,$\approx$\,0.13\,Jy (Fig.\,\ref{fig:spec_with_nebular}) and a standard optically-thick synchrotron spectral slope fall-off $\alpha$\,=\,--0.7 ($F_\nu$\,$\propto$\,$\nu^{\alpha}$) yields $B$\,$\ge$\,2\,$\times$\,10$^{4}$\,G and a zone radius $R$\,$\le$\,2\,$\times$\,10$^9$\,cm. Similar values have been inferred at the jet base of other systems \citep[e.g., ][]{Gandhi-2011, chaty11,Russell2014,Russell2020, echiburur24}, but the lack of secure detection of \nub\ precludes more detailed insight and comparison. Similarly to the irradiated disc scenario, our measurement of a weak r.m.s. from the timing data requires the compact jet to be highly stable in flux, which may be difficult to explain given the inferred radius of the jet base above. 

Data spanning a wider wavelength coverage will be required to test both the irradiated disc and the jet hypotheses.

\subsection{Bolometric Luminosity Estimates from the MIR}
\label{sec:lboldiscussion}

The emission line analysis above also allows approximate constraints on the ionising power of the intrinsic source ($L_{\rm Bol}$). This is because the Lyman continuum photon rate can be written as

\begin{equation}
    Q(H^0)=\Omega_{\rm ion}\int_{\nu_0}^{\infty}{\frac{L_\nu}{h\nu}d\nu}
    \label{eq:ionisationluminosity}
\end{equation}

\noindent
where $E_0$\,=\,$h\nu_0$\,=\,13.6\,eV, and $\Omega_{\rm ion}$ is the covering factor of the illuminated emission line gas. If a fraction $\Omega_{\rm w}$ of the sky as seen from the nucleus is obscured by optically thick matter which blocks the escape of ionising radiation, then $\Omega_{\rm ion}$\,=\,$1-\Omega_{\rm w}$. 

The intrinsic spectral shape is unknown, especially in the current obscured state of the source. But using Eq.\,\ref{eq:ionisationluminosity} above as a `bolometric indicator,' it is possible to constrain the source power approximately under different assumptions of the spectral shape. We illustrate this through two plausible example spectral shapes below. 

Using the estimate of $Q(H^0)$ from \S\,\ref{sec:linefluxconstraints} and assuming a simple power-law model with X-ray photon-index $\Gamma$\,=\,2 ($N_E$\,$\propto$\,$E^{-\Gamma}$, or $L_\nu$\,$\propto$\,$\nu^{-\Gamma+1}$ in Eq.\,\ref{eq:ionisationluminosity} above) extending over 0.013--100\,keV (say), we find $L_{\rm Bol}$\,$\sim$\,6\,$\times$\,10$^{37}$\,erg\,s$^{-1}$ where $L_{\rm Bol}$ is taken to closely approximate the ionising source power. Alternatively, if a multicolour accretion disc \citep{mitsuda84} with an inner temperature of $\approx$\,1\,keV  dominates the intrinsic UV--to--X-ray spectrum (cf.\,\citealt{rahoui10}), we require $L_{\rm Bol}$\,$\sim$\,5\,$\times$\,10$^{38}$\,erg\,s$^{-1}$.

The estimates above provides some first insight on $L_{\rm Bol}$, as derived from the MIR in the obscured state of \grs. Taken at face value, these imply that the intrinsic accretion power could plausibly range over $\approx$\,5--30\%\ of $L_{\rm Edd}$, though there are several important uncertainties to keep in mind.

Firstly, simulations treating the photoionisation around the source self-consistently need to be carried out in order to test the viability of the above estimates. A full exploration of the model parameter space -- which is highly degenerate in terms of geometry, optical depth, local physical conditions and the radiation spectrum of the central engine -- is well beyond the scope of this work. But we did conduct preliminary tests using the photoionisation code {\tt CLOUDY} \citep{cloudy98, cloudy23}, by assuming blackbody as well as power-law central ionising sources with appropriate ionising photon-rate estimates illuminating spherical or slab gas geometries with density estimates as derived above. Our main finding was that while both a blackbody or a power-law ionising source can reproduce the flux ratios of the primary emission lines, it is difficult to simultaneously produce the high observed {\em fluxes} of the emission lines. Most of our tests fell short in predicted line strength, and required the introduction of additional gas on scales extending to $r$\,$\gtsim$\,1,000\,lt-sec or more to reproduce the observed values. If not a numerical artefact, such a physically extended medium may be explained as the left-over ejecta that have accumulated in the circumbinary medium over time, and could also account for the relatively cool temperatures inferred for the MIR emission line region (cf.\,Fig.\,\ref{fig:spec_with_nebular}). However, we caution that convergent solutions were non-trivial to find and often sensitive to the boundary physical conditions -- a sign that optical depth effects at the high densities being simulated likely play a non-trivial role, unsurprisingly. We also found that harder ionising spectra tended to overproduce the fluxes of \heii\ (and other high excitation) lines which we do not observe, with softer spectra instead being preferred.

Another uncertainty is the covering factor $\Omega_{\rm ion}$ of the illuminated emission line gas, which remains unknown. If most lines-of-sight around the source are heavily obscured, with only a small fraction of the inherent ionising power escaping out, then $\Omega_{\rm ion}$ will be small and  $L_{\rm bol}$ will need to be correspondingly boosted upwards. This could be tested by searching for signatures of waste heat emerging in the far-IR or sub-mm regimes.

With the above caveats in the mind, the range of \lbol\ values herein shows that \grs\ is at least consistent with the central engine accreting at a moderately high, but sub-Eddington, accretion rate in its present X-ray--obscured state. Detailed validation will be required in future work to test this.

\subsection{The Origin of the MIR Emission of GRS 1915+105}
\label{sec:finaldiscussion}

The unique `X-ray--obscured' state of \grs\ has now lasted for several years. Substantial {\em radio} flaring has been reported as the X-rays have faded systematically over this time \citep{motta21}. Here, we have presented data demonstrating clear {\em MIR} brightening and flaring, both in terms of the long-term MIR monitoring shown in Fig.\,\ref{fig:longterm}, as well as our pointed \jwst\ observation which caught the source near the peak of one of these long-term MIR-flaring periods. This adds to evidence pointing to the presence of of an intrinsically active central engine, despite its apparent weakness in X-rays. 

The dereddened MIRI continuum luminosity of 10$^{36}$\,erg\,s$^{-1}$ is about $7\times10^{-4}$\ of the Eddington power (\S\,\ref{sec:powerdiscussion}). Using the emission lines as probes of the intrinsic ionising power, together with plausible assumptions on the shape of the intrinsic continuum suggests a bolometric (ionising) accretion luminosity $L_{\rm Bol}$\,$\approx$\,5--30\,\% $L_{\rm Edd}$ (\S\,\ref{sec:lboldiscussion}). This is subject to much uncertainty, not only on the spectral shape, but also on the covering factor of the illuminated emission line nebula, and high \lbol\ values cannot be ruled out. Nevertheless, such a range is qualitatively consistent with inferences from detailed X-ray studies that the source is  accreting at sub-Eddington rates \citep{miller20, balakrishnan21}. In any scenario where the MIR continuum is dominated by reprocessing of the intrinsic accretion power, the efficiency of reprocessing must then be high: $L_{\rm MIRI}$/\lbol\ ranges over 0.002--0.02 for the bolometric luminosity range above. When comparing to historical observations with \spitzer\ (Fig.\,\ref{fig:spitzer}), this ratio is now about an order-of-magnitude higher, presumably as a result of enhanced reprocessing in the circumnuclear gas which now persistently obscures the central source. 

We find a rich MIRI recombination line spectrum, which is also much brighter than seen in historical MIR spectroscopy before the source entered the obscured state. The MIRI spectrum can be approximated as emission from an optically-thin shell with mean electron temperature $T$\,$<$\,20,000\,K (and possibly close to 16,000\,K) and gas density \nh\,$\sim$\,10$^{10}$\,cm$^{-3}$ (\S\,\ref{sec:linediscussion}). The fact that most of the emission lines are less variable (\S,\ref{sec:timinglagresults}) than the continuum is consistent with physically segregated origins of the two. The only exceptions are the two \heii\ lines [(9--8)$\lambda$6.947\,\micron\ and (10--9)$\lambda$9.712\,\micron; cf. Fig.\,\ref{fig:rmscompare}], which likely arise closer in to the source of the continuum if helium is to be photoionised. 

At first glance, one might interpret the long emission line lag as the light-travel time to a physically extended gas shell on size scales of $\sim$\,few hundred lt-sec. However, given the relatively high densities inferred from our analysis (\S\,\ref{sec:densitydiscussion}), it is worth estimating the mean recombination time ($\tau_{\rm rec}$) in the emission line region. This can be estimated as follows:

\begin{equation}
    \tau_{\rm rec} = \frac{1}{n_{\rm H}\alpha_{\rm B}}
\end{equation}

\noindent
For the range of \nh\ values ($\ltsim$\,10$^{10}$ to $\approx$4\,$\times$\,10$^{10}$\,cm$^{-3}$) estimated in the emission line region (\S\,\ref{sec:densitydiscussion}), $\tau_{\rm rec}$ ranges from $\approx$\,100\,s to $\gtsim$\,400\,s. 

These timescales are similar to those inferred from the emission line lag analysis, and could partly account for the degeneracy that we inferred between the mean lag $\tau$ and smearing length $\sigma$ when using the Gaussian transfer function (\S\,\ref{sec:timingdiscussion}). Radiative transfer effects within dense media could add additional delays. In other words, the lag that we find should not be interpreted solely in terms of a light-travel time delay. It should also be kept in mind that the line response is non-stationary, with the lag itself changing toward the end of our observation (\S\,\ref{sec:timinglagresults}). Restricting our transfer function analysis to the initial 5,000\,s of the observation results in a characteristic lag 
$\tau$\,=\,6$_{-4}^{+8}$\,$\times$\,10$^1$\,s and smearing parameter $\sigma$\,=\,5.4\,(\p 0.5)\,$\times$\,10$^2$\,s. These timescales are still consistent with the need for a long lag and/or smearing of the emission line as inferred from the full observation (\S\,\ref{sec:results}), though longer duration sampling is needed to quantify these results more accurately. 

The \spitzer\ MIR spectra were proposed to span a transition (occurring at $\approx$\,10\,\micron) between the outer irradiated disc and a cold ($T$\,$\sim$\,300--500\,K) dust component heated by the companion star \citep{rahoui10}. Our results are not easily consistent with such an interpretation, for two reasons: (1) The shape of the observed MIRI spectrum is remarkably similar to that seen by IRS, despite the source having brightened by a factor of 7--13 in the interim (Fig.\,\,\ref{fig:spitzer}). Preserving the spectral shape over the years would require fine-tuning to scale up both components identically; (2) If the cold dust component were heated by the companion star, its luminosity would not be expected to differ between the \spitzer\ and \jwst\ epochs, leading to a change in spectral shape across the transition, which we do not obviously see. The apparent lack of PAH features in our data (\S\,\ref{sec:dustdiscussion}) argues for the dust component to have been heated, and possibly destroyed in the intervening $\approx$\,17 years since their detection, by a change in ionising conditions, which also argues for heating by the compact object instead of the companion star. 

In fact, it is non-trivial to preserve the spectral shape in most multi-component physical scenarios in which the source flux changes dramatically. Instead, a (single-component) power-law could achieve this, and the dereddened MIRI continuum (Fig.\,\ref{fig:spitzer}) can also be qualitatively explained as an approximately flat power-law between 5--10\,\micron\ (with a slight rise at the blue end below the \spitzer\ IRS short wavelength limit, possibly signalling a second component). The source has also shown enhanced radio activity since the 2018 obscured state \citep{motta21}, though we cannot say whether any radio power-law synchrotron component extends from the radio to the MIR herein. One caveat regarding the modelling of the continuum shape is the apparent spectral calibration uncertainty at both ends of the MIRI wavelength range. Whilst we have attempted to correct for these (Appendix\,\ref{sec:1033correction}), this needs to be verified and cross-checked against other multiwavelength data. 

Interpreting the MIRI continuum in terms of thermal bremsstrahlung from the obscuring wind requires a wind density of \nh\,$\sim$\,10$^{11}$\,cm$^{-3}$, about one order of magnitude denser than the emission line region (\S\,\ref{sec:densitydiscussion}). The inferred wind density is also similar to the gas density inferred via X-ray absorption, if this absorbing gas column spans the outer accretion disc. So the wind could potentially explain the MIRI continuum, the MIR emission lines (as the outer optically-thin fringes of the wind), and could also be the same medium that obscures the X-rays. 

While the above scenario may be one self-consistent interpretation of the data, it pushes us into an extreme regime in terms of the strength of matter feedback. At the mass loss rate required by the wind  (\S\,\ref{sec:mass}), the disc would lose a solar mass in $\approx$\,10$^4$\,y, requiring full replenishment every year (based upon the critical estimates of the disc mass; \citealt{koljonen21_grs}). In terms of the Eddington accretion rate $\dot{M}_{\rm Edd}$\,=\,$L_{\rm Edd}/\eta c^2$\,=\,2.6\,$\times$\,10$^{-7}$\,\Msun\,yr$^{-1}$ with standard radiative efficiency $\eta$\,=\,0.1, the source is losing mass at $\approx$\,10$^3$\,$\dot{M}_{\rm Edd}$. This would clearly be an unsustainable rate for the source to maintain over a long period of time. 

The mass-loss could be moderated if the wind speed was itself variable and this is, indeed, observed. An irregular {\em NIR} wind has been observed on several occasions, with speeds exceeding $10^3$\,km\,s$^{-1}$ \citep{sanchezsierras23}. At these speeds, though, the wind cannot remain bound, so such episodes must be transient. Our \jwst\ observations were conducted very soon after the peak of unprecedented MIR and radio flaring activity (cf.\,Fig.\,\ref{fig:longterm}), with the source approaching and exceeding radio flux densities of $\approx$\,1\,Jy over Apr--Aug\,2023 \citep{trushkin23, egron23, bright23_atel}. In other words, the physical parameters that we infer could simply be reflecting the short-lived, albeit extreme, nature of the source state during the past few years. Future MIR observations at more moderate flux levels should be able to easily test this hypothesis. 

By contrast, the wind seen in {\em X-rays} originates closer in to the nucleus and was seen to be an order-of-magnitude slower at the onset of the obscured state \citep{miller20}. In fact, this hot wind may well be a `failed outflow' which does not escape to large scales. If the MIR bremsstrahlung continuum is probing the remnants of such a failed wind, our inferred mass-loss rates would be substantially lowered. Understanding the interplay between the multiple phases of such a wind should be possible through joint X-ray microcalorimeter and MIR high-spectral resolution observations. We also note that we cannot rule out the alternative possibilities of jet or irradiated accretion disc contributions to the MIR continuum (though these also have associated difficulties; \S\,\ref{sec:othercontinuum}), nor have we investigated shock ionisation as a mechanism to produce the emission lines observed. Under any of these scenarios, the mass-loss rate would naturally come down, but alternative mechanisms would be required to suppress the expected MIR emission from the excess gas that must be obscuring the source in X-rays.

It is also worth questioning whether \grs\ is unique amongst the XRB population in terms of the extreme behaviour described above, and if so, why. The presence of extreme obscuration has been inferred in other well-studied XRBs, e.g. V404\,Cyg \citep{motta17, walton17} and Swift\,J1357.2-0933
 \citep{charles19}. In the former case, extreme obscuration episodes were short-lived, with the wind clearing the circumnuclear medium in a matter of weeks to months. \grs\ has a much larger accretion disc than both these systems \citep[and, in fact, has the largest disc of all known Galactic XRBs;][]{casaresjonker14}, affording a much larger reservoir to accumulate gas. The inclination angle of \grs\ ($i$\,$\approx$\,64\,deg) is also relatively high, so it is easier for gas near the disc plane to cause obscuration along the line-of-sight (as is thought to be the case in Swift\,J1357.2-0933). Pinning down the fundamental cause of the transition that resulted in the source entering its current X-ray--obscured state will be key to understanding its relation to other XRBs. \grs\ continues to be a fascinating source for studies of accretion, even in its current exceptional state.
 
\section{Conclusions}

\grs\ has brightened dramatically in the MIR compared to historical averages. The now-persistent (X-ray) obscured state is thus characterised by MIR brightening and long-term flaring, in addition to the activity that has been reported in the radio.

MIRI also reveals the presence of a rich emission line spectrum. Some of the strongest recombination emission lines may have been present in previous \spitzer\ IRS observations, but only at much weaker S/N, and may have been variable and/or confused with adjacent features \citep{rahoui10, harrison14}. Detailed timing analysis in our MIRI data now shows indications of an emission line lag relative to the underlying continuum. The lag is consistent with timescales characteristic of the outer accretion disc, and the relatively cool characteristic temperature ($T$\,$\ltsim$\,20,000\,K) suggests that the emission lines cannot originate very close to the source. However, the lag is also similar to the expected recombination time for its modelled gas density, so cannot be interpreted solely as a light-travel time delay. PAHs apparent in previous data have either remained at a similar strength or may have been destroyed in the interim. 

The source is thought to be obscured in X-rays by an optically-thick wind. The MIRI continuum could represent optically-thick bremsstrahlung emission from such a wind, with the emission lines arising from its outer nebular fringes. This is also commensurate with the integrated column densities inferred from X-ray continuum absorption, and self-consistently explains the lack of any strong flux variations which would have been washed out in propagating through the optically-thick medium. 

Using the flux of the most significant MIRI Pf$\alpha$ emission line as a bolometric indicator implies a moderate intrinsic accretion luminosity, though self-consistent photoionisation modelling in optically-thick media is needed to better constrain this. We are able to constrain the mass of the obscuring medium, but if estimates at the high end of wind speeds are adopted, this would imply an unsustainable rate of mass loss from the system. \jwst\ caught the source during a period of exceptional flaring during 2023, so these extreme inferences need not necessarily apply in the longer term. Irrespective, our analysis adds to evidence arguing that \grs\ remains active during the `obscured' state, and is perhaps growingly erratic in its multiwavelength behaviour now. 

These results not only give new insight into the nature of the unique obscured state of \grs, but also showcase possibilities of rapid MIR spectral-timing analysis. \jwst's unparalleled sensitivity was crucial for uncovering the weak ($\sim$\,1\%) , but highly significant r.m.s. that we observe, and for detection of the emission line lag. We have also highlighted systematic uncertainties that impact current observations in the MIRI LRS spectral-timing mode, and which need to be accounted for when searching subtle features. \jwst\ is still a young mission, and better understanding of these systematic issues is expected as knowledge of the MIRI calibration and the reduction pipeline improve with time. 

Further work is ongoing to collate and analyse the multiwavelength spectral energy distribution of \grs\ which should help to test some of the questions raised herein, e.g. on the possibility of jet or irradiated accretion disc contributions to the MIR. Future observations at higher spectral resolution (with \jwst's MRS instrument; \citealt{jwstmrs}) will allow superior line deconvolution and searches for wind signatures in the MIR.

\section*{Data Availability}

The core data analysed herein are publicly available in telescope archives. The \jwst\ data may be found using the Program identifiers  1586 and 1033. The NEOWISE, MAXI and {\em RXTE} data are similarly publicly available. AMI data can be made available upon reasonable request to the coauthors.

\section*{Acknowledgments}

We thank the mission and instrument teams at STScI for their patience and expert help with our numerous queries, especially MIRI scientist S. Kendrew and staff astronomer I. Wong. 

PG acknowledges funding from The Royal Society (SRF$\backslash$R1$\backslash$241074). PG and MR thank UKRI Science \& Technology Facilities Council for support. AJT acknowledges the support of the Natural Sciences and Engineering Research Council of Canada (NSERC; funding reference number RGPIN-2024-04458). TJM acknowledges support from JWST-GO-01586.002. RIH and ESB acknowledge support from JWST-GO-01586.007. JAT acknowledges support from JWST-GO-01586.010-A.
TS and FMV acknowledge financial support from the Spanish Ministry of Science, Innovation and Universities (MICIU) under grant PID2020-114822GB-I00. JAP acknowledges support from STFC consolidated grant ST/X001075/1.
RMP acknowledges support from NASA under award No. 80NSSC23M0104.
DP acknowledges the support from ISRO (India), under the ISRO RESPOND program. The work of MER was carried out at the Jet Propulsion Laboratory, California Institute of Technology, under a contract with the National Aeronautics and Space Administration. MCB and TDR acknowledge support from the INAF-Astrofit fellowship. GRS and COH are supported by NSERC Discovery Grants RGPIN-2021-0400 and RGPIN-2023-04264, respectively. DMR is supported by Tamkeen under the NYU Abu Dhabi Research Institute grant CASS. P.S-S. acknowledges financial support from the Spanish I+D+i Project PID2022-139555NB-I00 (TNO-JWST) and the Severo Ochoa Grant CEX2021-001131-S, both funded by MCIN/AEI/10.13039/501100011033. VSD acknowledges support by the Science and Technology Facilities Council (grant ST/V000853/1). SM is supported by a European Research Council (ERC) Synergy Grant "BlackHolistic" grant No. 10107164.

Line identification benefited from the compilation v3.00b4 presented in \citet{linelist}.\footnote{\url{https://linelist.pa.uky.edu/newpage/}} Cloudy calculations were performed with version c23.01 \citep{cloudy23}.

This work is based on observations made with the NASA/ESA/CSA {\em James Webb Space Telescope}. The data were obtained from the Mikulski Archive for Space Telescopes at the Space Telescope Science Institute, which is operated by the Association of Universities for Research in Astronomy, Inc., under NASA contract NAS 5-03127 for JWST. These observations are associated with program \#1586 \citep{jwst1586}. Support for program \#1586 was provided by NASA through a grant from the Space Telescope Science Institute, which is operated by the Association of Universities for Research in Astronomy, Inc., under NASA contract NAS 5-03127.

This research has made use of the NASA/IPAC Infrared Science Archive, which is funded by the National Aeronautics and Space Administration and operated by the California Institute of Technology. The DOI of the \spitzer\ Enhanced IRS Products is 10.26131/IRSA399. 

This research has made use of MAXI data provided by RIKEN, JAXA and the MAXI team.

This research has made use of data and/or software provided by the High Energy Astrophysics Science Archive Research Center (HEASARC), which is a service of the Astrophysics Science Division at NASA/GSFC.

Based on observations with ISO, an ESA project with instruments funded by ESA Member States (especially the PI countries: France, Germany, the Netherlands and the United Kingdom) and with the participation of ISAS and NASA.

PG is grateful to S.\,F.\,H\"{o}nig, J.\,Hernandez-Santisteban and J.\,H.\,Matthews for discussions at the initial stages of analysis. 

\newpage

\newpage
\bibliography{GRS1}
\bibliographystyle{mnras}

\appendix

\section{Comparison with archival stellar observations in the SLITLESS PRISM Time Series mode}
\label{sec:1033correction}

During early commissioning, an observation of the bright star, L168--9b, was taken in MIRI SLITLESS PRISM mode in order to calibrate LRS time series observations (TSO; \citealt{bouwman23}). L168--9b is known to host a transiting exoplanet, but is otherwise an ordinary M1V star. It is also approximately similar in MIR brightness to \grs\ at the time of our observation. This commissioning observation thus provides a good comparator in order to assess instrumental systematics. 

The commissioning data (Programme ID 1033) were downloaded, reduced and extracted through the \jwst\ pipeline using parameters identical to those adopted for \grs. The TSO data in this case comprise 9,371 on-sky integrations, each 1.431\,s in duration. Detailed analysis of these data can be found in \citet{bouwman23} and \citet{dyrek24}. 

\begin{figure}
    \centering
    \includegraphics[width=\columnwidth]{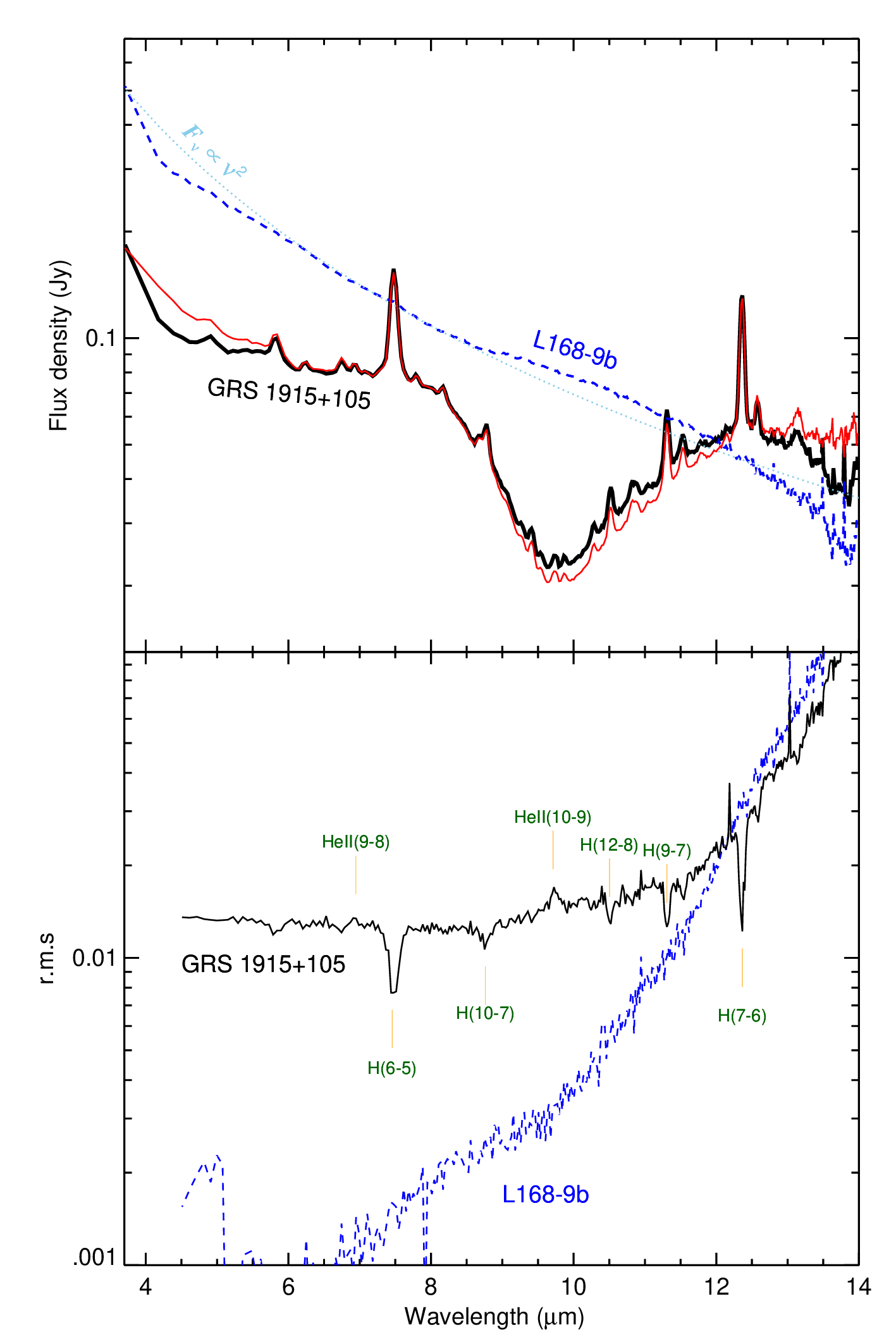}
    \caption{Median spectrum {\em (Top)} and excess r.m.s. {\em (Bottom)} of \grs\ compared to archival data of the commissioning observation of star L168--9b (blue dashed). In the top panel, the dotted light-blue line shows an expected Rayleigh--Jeans tail ($F_\nu$\,$\propto$\,$\nu^2$) normalised to L168--9b at 8\,\micron, used to determine the residual spectral calibration corrections. The resultant corrected spectrum of \grs\ is shown in red. In the bottom panel, the wavelengths of prominent emission lines are annotated.}
    \label{fig:rmscompare}
\end{figure}

The median flux-calibrated spectrum of L168--9b is shown in Fig.\,\ref{fig:rmscompare}, together with the spectrum of \grs. Some common trends in the continua of both objects are immediately apparent, specifically at the two ends of the wavelength range -- at short wavelengths, both spectra have a convex shape between $\approx$\,4--5\,\micron\ with a steep rise below 4\,\micron, while the shape is concave beyond $\approx$\,12\,\micron\ where it curves downwards with identical narrow features in both between $\approx$\,13.5--14\,\micron. The absolute flux calibration beyond $\approx$\,12\,\micron\ is known to be uncertain \citep[e.g., ][]{kendrew15}, while the shortest wavelength bin at $\approx$\,3.7\,\micron\ lies outside the formal MIRI wavelength range. The fact that similar continuum features are seen in both spectra confirms that they must be artefacts of imperfect spectral calibration. 

There is a known spectral fold-over at the shortest wavelengths arising from reflection and scattering contamination \citep{miricommissioning}, so we do not utilise data below 4.5\,\micron\ for our science. Above this wavelength, we attempt to correct for the aforementioned imperfections by comparing the MIRI spectrum of L168--9b to a Rayleigh-Jeans (RJ) tail of a blackbody. The flux density in such a tail should scale as $F_\nu$\,$\propto$\,$\nu^2$. As seen in Fig.\,\ref{fig:rmscompare}, L168--9b follows the expected scaling well between $\approx$\,6--8.5\,\micron. By normalising at 8\,\micron, we are thus able to correct departures from the RJ tail at both wavelength ends (and all other wavelengths). The assumption here is that the star can be described by RJ. This is not unreasonable for the MIRI range, given that the spectral peak for an M1 star (with effective temperature $T_{\rm eff}$\,$\approx$\,3600\,K) will lie at a much shorter wavelength around 0.8\,\micron. But it does not account for any intrinsic departures, e.g. due to the presence of any dust in the system, and should thus be treated with some caution for precision absolute flux measurements. 

With this caveat in mind, applying the RJ correction to our observed spectrum of \grs\ does result in a smoother continuum at both ends (Fig.\,\ref{fig:rmscompare}). Measurement of a more accurate spectral shape will require longer wavelength observations with MIRI/MRS in future cycles, and comparison with NIR spectroscopy which is beyond the scope of this work. We adopt this correction for our science spectrum presented herein. 

The bottom panel of Fig.\,\ref{fig:rmscompare} shows the comparison of the r.m.s. computed according to Eq.\,\ref{eq:rms} for \grs\ and L168--9b. The latter star is not expected to be variable in the MIR (except for small systematic flux deviations caused by the exoplanet transit). The rise to the MIR seen in both systems is clearly an indication of systematics and we thus cannot trust the red continuum r.m.s. rise in \grs. A similar discussion can be found in \citet{dyrek24}, where the rise in r.m.s is attributed to a diminishing S/N in a fixed-width extraction aperture at long wavelengths as the spectrum starts to become increasing background-dominated. For \grs, the surface brightness of the background exceeds 50\,\% of the source at $\lambda$\,$\gtsim$\,12.5\,micron, and grows to about 200\,\% relative to the source at the longest wavelength of 14\,\micron. Uncertainties related to known read-noise underestimation could additionally contribute here. 

Other than this, L168--9b reassuringly shows much smaller r.m.s. at short wavelengths, implying that our detection of flux variations over the bulk of the MIRI range in \grs\ is robust. 

\section{Theoretical H and He wavelengths}
\label{sec:hseries}

Table\,\ref{tab:Hseries} lists the Rydberg series wavelengths of prominent H and \heii\ transitions in the MIRI range. These are listed as a reference to facilitate comparison with the observed spectra. 

\begin{table}
    \centering
    \caption{Theoretical wavelengths of some of the most prominent Hydrogen Series and Helium\,II emission lines in the MIRI wavelength range.}
\begin{tabular}{l}
    \hline
    \hline
    Feature \hspace{1.2cm} Wavelength\\
             \hspace{2.99cm} \micron\\
    \hline
Pf\,(7--5) \dotfill\  4.6538 \\
Pf\,(6--5) \dotfill\  7.4599 \\
\\
Hp\,(11--6) \dotfill\  4.6725 \\
Hp\,(10--6) \dotfill\  5.1286 \\
Hp\,(9--6) \dotfill\  5.9082 \\
Hp\,(8--6) \dotfill\  7.5025 \\
Hp\,(7--6) \dotfill\ 12.3719 \\
\\
16--7 \dotfill\  5.5252 \\
15--7 \dotfill\  5.7115 \\
14--7 \dotfill\  5.9568 \\
13--7 \dotfill\  6.2919 \\
12--7 \dotfill\  6.7720 \\
11--7 \dotfill\  7.5081 \\
10--7 \dotfill\  8.7600 \\
9--7 \dotfill\ 11.3087 \\
\\
16--8 \dotfill\  7.7804 \\
15--8 \dotfill\  8.1549 \\
14--8 \dotfill\  8.6645 \\
13--8 \dotfill\  9.3920 \\
12--8 \dotfill\ 10.5035 \\
11--8 \dotfill\ 12.3871 \\
\\
17--9 \dotfill\ 10.2612 \\
16--9 \dotfill\ 10.8036 \\
15--9 \dotfill\ 11.5395 \\
14--9 \dotfill\ 12.5870 \\
\\
20--10 \dotfill\ 12.1568 \\
19--10 \dotfill\ 12.6109 \\
18--10 \dotfill\ 13.1880 \\
17--10 \dotfill\ 13.9417 \\
                    \\
    He\,{\sc ii}\,(8--7) \dotfill\ 4.7635 \\
    He\,{\sc ii}\,(9--8) \dotfill\ 6.9480 \\
    He\,{\sc ii}\,(10--9) \dotfill\ 9.7135 \\
    He\,{\sc ii}\,(11--10) \dotfill\ 13.1283 \\
    \hline
\end{tabular}
     \label{tab:Hseries}
 \end{table}

\section{Extinction curve for GRS 1915+105}
\label{sec:extinction}

Table\,\ref{tab:extinction} lists the values of extinction $A_\lambda$ spanning the optical to MIR towards \grs\ used herein. This assumes an \av\,=\,19.6\,\p\,1.7\,mag \citep{chapuis04}, combined with the infrared extinction curves of \citet{fitzpatrick99} and \citet{chiartielens06}. 

\begin{table}
    \centering
    \caption{Multiwavelength Extinction toward \grs}
\begin{tabular}{l}
    \hline
    \hline
    $\lambda$ \hspace{1.3cm} $A_{\lambda}$ \\
    \micron \hspace{1cm} mag \\
    \hline
    0.55 \dotfill\ 19.60 \\
    1 \dotfill\ 7.99\\
    2 \dotfill\ 2.41\\
    3 \dotfill\ 1.39\\
    4 \dotfill\ 1.06\\
    5 \dotfill\ 0.91\\
    6 \dotfill\ 0.84\\
    7 \dotfill\ 0.82\\
    8 \dotfill\ 0.84\\
    9 \dotfill\ 1.52\\
    10 \dotfill\ 2.12\\
    11 \dotfill\ 1.66\\
    12 \dotfill\ 1.31\\
    13 \dotfill\ 1.14\\
    14 \dotfill\ 1.16\\
    15 \dotfill\ 1.22\\
    16 \dotfill\ 1.29\\
    17 \dotfill\ 1.38\\
    18 \dotfill\ 1.47\\
    19 \dotfill\ 1.52\\
    20 \dotfill\ 1.48\\
    21 \dotfill\ 1.40\\
    22 \dotfill\ 1.31\\
    23 \dotfill\ 1.24\\
    24 \dotfill\ 1.16\\
    25 \dotfill\ 1.09\\
    26 \dotfill\ 1.03\\
    27 \dotfill\ 0.97\\
    \hline
    \hline
\end{tabular}
    \label{tab:extinction}
\end{table}

\section*{Affiliations}
{\small 
$^{1}$School of Physics \& Astronomy, University of Southampton, Southampton SO17 1BJ, UK\\
$^{2}$Department of Physics \& Astronomy, Louisiana State University, 202 Nicholson Hall, Baton Rouge, LA 70803, USA\\
$^{3}$Department of Physics \& Astronomy, Texas Tech University, Box 41051,Lubbock, TX 79409-1051, USA\\
$^{4}$Department of Physics and Astronomy, Butler University, 4600 Sunset Avenue, Indianapolis, IN 46208, USA\\
$^{5}$Cahill Center for Astronomy \& Astrophysics, California Institute of Technology, Pasadena, CA 91125, USA \\
$^{6}$INAF-Osservatorio Astronomico di Brera, Via Bianchi 46, I-23807 Merate (LC), Italy\\
$^{7}$Department of Astronomy and Astrophysics, Tata Institute of Fundamental Research, 1 Homi Bhabha Road, Colaba, Mumbai 400005, India\\
$^{8}$Space Telescope Science Institute, 3700 San Martin Drive, Baltimore, Maryland 21218, USA\\
$^{9}$South African Astronomical Observatory, P.O.Box 9, Observatory, 7935, South Africa\\
$^{10}$Independent\\
$^{11}$INAF-Osservatorio Astronomico di Roma, via Frascati 33, I-00078 Monteporzio Catone (RM), Italy\\
$^{12}$Department of Physics, University of Warwick, Gibbet Hill Road, Coventry CV4 7AL, UK\\
$^{13}$European Southern Observatory, Alonso de C\'ordova 3107, Casilla 19001, Vitacura, Santiago, Chile, Coventry CV4 7AL, UK\\
$^{14}$Department of Physics and Astronomy, University of Sheffield, Sheffield, S3 7RH, UK\\
$^{15}$Instituto de Astrof\'{i}sica de Canarias, E-38205 La Laguna, Tenerife, Spain\\
$^{16}$Astrophysics, Department of Physics, University of Oxford, Keble Road, Oxford OX1 3RH, UK\\
$^{17}$Department of Physics, University of Alberta, CCIS 4-181, Edmonton, AB T6G 2E1, Canada \\
$^{18}$ Anton Pannekoek Institute for Astronomy \& Gravitation Astroparticle Physics Amsterdam (GRAPPA) Institute, University of Amsterdam, Science Park 904, 1098XH Amsterdam, The Netherlands\\
$^{19}$ Dipartimento di Fisica, Universit\`{a} degli Studi di Milano, Via Celoria 16, I-20133 Milano, Italy\\
$^{20}$Harvard-Smithsonian Center for Astrophysics, 60 Garden St., Cambridge, MA, USA\\
$^{21}$Department of Astronomy, University of Michigan, 1085 South University Avenue, Ann Arbor, MI 48109, USA\\
$^{22}$ICRAR -- Curtin University, GPO Box U1987, Perth, WA 6845, Australia\\
$^{23}$INAF-Osservatorio Astronomico di Brera, Via Bianchi 46, I-23807 Merate (LC), Italy\\
$^{24}$Centre for Extragalactic Astronomy, Department of Physics, Durham University, South Road, Durham DH1 3LE, UK\\
$^{25}$Department of Physics, R. J. College, Mumbai 400086, India \\
$^{26}$Department of Physics, University of Nevada, Reno, NV 89557, USA \\
$^{27}$Department of Physics \& Astronomy, Embry-Riddle Aeronautical University, 3700 Willow Creek Road, Prescott, AZ, 86301, USA\\
$^{28}$Jet Propulsion Laboratory, California Institute of Technology, 4800 Oak Grove Drive, Pasadena, CA 91109, USA\\
$^{29}$Center for Astrophysics and Space Science (CASS), New York University Abu Dhabi, PO Box 129188, Abu Dhabi, UAE\\
$^{30}$INAF - IASF Palermo, via Ugo La Malfa, 153, I-90146 Palermo, Italy\\
$^{31}$Instituto de Astrof\'{\i}sica de Andaluc\'{\i}a (CSIC), Glorieta de la Astronom\'{\i}a s/n, 18008-Granada, Spain\\
$^{32}$Departamento de Astrof\'{ı}sica, Universidad de La Laguna, E-38206 La Laguna, Tenerife, Spain\\
$^{33}$Department of Physics and Astronomy, University of Lethbridge, Lethbridge, Alberta, T1K 3M4, Canada\\
$^{34}$Space Sciences Laboratory, University of California, 7 Gauss Way, Berkeley, CA 94720-7450, USA\\
\\
}
\bsp	
\label{lastpage}
\end{document}